\documentclass[aps,twocolumn,10pt,pre]{revtex4-1}
\usepackage{graphicx} 
\usepackage{multirow} 
\usepackage{longtable, tabu} 
\usepackage{hhline}
\newtabulinestyle{mydashline=on 1.5pt off 2pt}
\usepackage{amsmath,amsfonts, bm}
\usepackage{tikz}

\begin{document}
\title{Graph conductance, synchronization, and a new local bottleneck measure for decentralized network optimization}
\begin{abstract}
The two most prominent bottleneck measures in graph theory, commonly known as the isoperimetric number and conductance, are both referred to in the literature as Cheeger constants, which is an eponymous term originating from Jeff Cheeger's work in Riemannian geometry.  While these measures are useful for assessing barriers to flow in networked systems, neither is sufficient to characterize the stability of complete synchronization, i.e. the dynamic convergence of every node in the system toward a synchronization manifold.  Conductance, in particular, does provide an effective bound on the coupling strength required to achieve bulk synchronization, but it often fails to differentiate between many chimera states. The Fiedler vector, which is the eigenvector associated with the algebraic connectivity of the graph, is similarly often able to correctly identify the limiting cut, however, it can fail as well being a global relaxation, especially in the presence of small-set bottlenecks. Furthermore, the computations of these properties are NP-Hard and require global information about the network structure, which is often unrealistic for many complex systems.  We define a normalized \textit{synchronization bottleneck ratio} for each bi-partitioning graph cut.  The minimum of this ratio over all such cuts gives the \textit{synchronization bottleneck measure} as a global network property, and the associated cut better identifies the true limiting bottleneck for complete synchrony for a specific class of network coupled dynamics. Obtaining this global minimizer is also NP-hard, but due to the use of only local information in the argument, heuristics based on the ratio itself can guide decentralized strategies for improving the synchronizability in man-made networked systems; having relevance in power grid stability, distributed database coherence, and other applications where the alignment of every node is required for the proper functioning of the system. 
\end{abstract}

\author{C. Tyler Diggans}
\affiliation{Root Dynamix: A Complex Systems Science Company, Sewickley, PA 15143}
\affiliation{Department of Electrical and Computer Engineering, Clarkson University, Potsdam, NY}
\author{Jeremie Fish}
\affiliation{Department of Electrical and Computer Engineering, Clarkson University, Potsdam, NY}
\author{Abd AlRahman R. AlMomani}
\affiliation{Department of Mathematics, Embry-Riddle Aeronautical University, Prescott, AZ}

\maketitle
    
\section{Introduction}
Synchronization in coupled dynamical systems has been widely studied for more than half a century under various circumstances~\cite{Kuramoto75,Pecora90,rosenblum1996phase,Nishikawa03,belykh2005synchronization,sun2009synchronization} and for a wide range of applications~\cite{Winfree67,ravoori2011robustness,motter2013spontaneous,Gallego17}; this system-level phenomenon increasingly appears relevant to the proper functioning of many emergent and distributed systems~\cite{boccaletti2006complex,dorfler2014synchronization}. As such, a large body of literature has been devoted to the optimization of network structures for more robust synchronization stability, e.g.,~\cite{restrepo2006characterizing,pecora2008synchronization,Hagberg08,jalili2013enhancing,skardal2014optimal,fazlyab2017optimal}, but existing network bottleneck measures fail to fully characterize \textit{complete} stability, wherein the dynamics of every node in the system converges toward a synchronization manifold.  This kind of system stability is important for many man-made systems such as power grids and distributed databases, where having even a single node far from synchrony can cause system failures.

Symmetric diffusive coupling serves not only to exchange information between adjacent node dynamics~\cite{bollt2012synchronization}, but ultimately to dissipate the entropy of the collective dynamic state over time.  Even when the node dynamics are chaotic, a stable synchronization manifold can result if the dissipation rate is such that the overall rate of entropy production within the system is properly bounded; this can enable convergence toward a manifold around a shared collective dynamic state, even when the node dynamics are not necessarily identical~\cite{sun2009master}.  Importantly, while the production of entropy is localized at the node dynamics, in the form of perturbations or local system updates, its dissipation occurs along the edges of the network structure and is controlled by the edge weights, introducing heterogeneity into the diffusive process on complex networks. 

The coupling network topology then becomes central for determining the stability for synchronization within a system.  Global stability analysis was explored for highly structured networks, e.g.,~\cite{belykh2000hierarchy,belykh2003lattice}, and an innovative approach was proposed in~\cite{belykh2004connection} for more general networks that focused on the collection of paths in a graph that utilize each edge.  This approach is related to the effective resistance network that follow Kirchhoff's laws and provides a robust bottleneck measure, itself being related to the Moore-Penrose Laplacian. But, for most applications, global stability is impractical and the requisite knowledge about the network is not available.  The more relevant question across man-made applications such as power grid dynamics is whether a system dynamics will be locally stable under small perturbations~\cite{motter2013spontaneous}, and will the system return to the synchronous state spontaneously and rapidly. 

Much of the synchronization literature has focused on this question for linear diffusive coupling as defined by the combinatorial Laplacian matrix, but we need not restrict our analysis to this case only.  Specifically, there are many applications for which the random walk Laplacian and/or the associated symmetric normalized Laplacian might be more relevant.  These are generally applications where each node has an internal dynamic state and a limited proportional capacity for incorporating information from its neighboring nodes.  Even simple clock disciplining often fits this description to an extent, where slewing is used to ensure monotonicity. As more man-made distributed computational systems are designed, this form of synchronization may become more relevant in practice.  In addition, the deep connection between these different Laplacian matrices and their associated bottleneck measures are central to our discussion and provide relevant context.   

The concept of a Master Stability Function (MSF) was introduced in~\cite{Pecora98}, enabling the analytical separation of the impact of the network topology from that of the particular dynamics and coupling functions on the local stability of synchronization for a large class of systems. This innovation resulted in an increased focus on spectral graph theory in the study of synchronization.  And, this was particularly true in the case of chaotic dynamics where positive topological entropy oscillators bring a more complex consideration of information generation, flow, and dissipation through diffusive coupling~\cite{huang2009generic}. 

At the heart of spectral graph theory are a pair of network properties that are often collectively referred to as Cheeger constants due to their origin being in analogy with the work of Jeff Cheeger on isoperimetric numbers for Riemannian manifolds~\cite{Cheeger70}.  Solving for either of these graph properties involves combinatorial optimization over the subsets of the vertex set, but they are each related to the smallest non-zero eigenvalue of an associated Laplacian matrix (both being discrete analogs of the Laplace–Beltrami operator~\cite{nadler2005diffusion}) through their associated Cheeger inequalities~\cite{mohar1989isoperimetric,Chung96}.

These well established spectral relations, along with the importance of the smallest non-zero eigenvalue in the MSF formalism, indicate that these bottleneck measures may have a role in determining the synchronizability of diffusively coupled systems~\cite{donetti2005entangled,huang2009generic,diggans2021essential}. However, the Cheeger inequalities generally provide loose bounds and the Fiedler vector, which is the eigenvector associated with the smallest non-zero eigenvalue~\cite{fiedler1973algebraic}, i.e. the algebraic connectivity, is defined by a continuous relaxation that favors balanced cuts over small-set bottlenecks.  As a result, these two well-known bottleneck measures and the common means of estimating the relevant cuts are more indicative of whether a majority of the nodes in a system will exhibit synchronization, what we will refer to as bulk synchronization.  

One additional approach to bottleneck identification is centered around computing the effective resistance for each link~\cite{dorfler2010synchronization}.  This often provides better estimates for small-set bottlenecks than the Fiedler vector due to a connection with the Moore-Penrose pseudoinverse, which we will discuss here.  This connection provides strong theoretical support, but it remains a globally focused approach that can be associated with the path counting heuristic and offers no guarantee either; though interesting decentralized approximation algorithms have recently been proposed~\cite{peng2021local}.  Regardless, all current methods fail to provide a full characterization of the specific critical links to local stability, and the ever-expanding instances of large man-made complex networked systems that lack global structural knowledge invite further inquiry into possible locally defined bottleneck measures related to complete synchronization.

Our goal is to identify the specific set of edges in a graph that will be the first to reach their collective capacity to transmit information between disjoint subsets of nodes during synchronization. Due to the wide range of systems that can be studied and the potential complexities in the process of synchronization, including coupling through uninformative variables, the actual limiting cut may not be identified, but our proposed measure provides a better approximation to the underlying bottleneck for synchronization than existing measures.  We intentionally limit our consideration here to a simplified case of identical dynamics under full variable coupling for clarity; while future work may explore multi-hop neighborhoods for different coupling functions and heterogeneous dynamics. 

A concise summary of the relevant background material is provided in Sec.~\ref{sec:background} including a review of the MSF formalism, and a summary of key relevant spectral graph theory results.  In Sec.~\ref{sec:Inter}, we then explore an application of the interlacing theorem that provides bounds for the algebraic connectivity in the special case of a light-weight pendant vertex attachment to an existing graph, which we use as a motivating example throughout.  

The synchronization bottleneck measure of a weighted symmetric graph is introduced in Sec.~\ref{sec:Xi} as the minimum of a normalized ratio.  We have chosen the symbol $\Xi_G$ to represent this bottleneck of a graph $G$ due to a combination of its limited use elsewhere in the graph literature and its symbolic connotation with its definition (see Fig.~\ref{fig:defns}).  

The desirable features of a synchronization bottleneck measure are validated in Sec.~\ref{sec:validate} by considering three specific examples,
and in Sec.~\ref{sec:verify}, we further validate the measure experimentally using information measures between node dynamics. 
We conclude this work with suggestions for decentralized strategies to improve synchronizability of man-made networked system based on our observations. 

\section{Background}
\label{sec:background}
To prevent the need for outside reference, here we provide the foundational concepts for stability analysis including the basic form of the Master Stability Function and a concise summary of the Cheeger constants, which form the basis for much of spectral graph theory. 
\subsection{Diffusively Coupled Dynamical Systems}
\label{sec:Sys}
Assume a collection of $N$ identical dynamical systems of the form $\dot{{\bf x}}=f({\bf x})$, with ${\bf x}\in\mathbb{R}^d$, diffusively coupled through a (weighted and undirected) network defined by the graph $G=(V,E,W)$, where $W$ is an $N\times N$ symmetric matrix consisting of positive real entries $W_{i,j}>0$ for each edge $\left\{i,j\right\}\in E$ and zero otherwise. Generalizations may apply to more complex coupling through consideration of their linearizations, but our focus will be on linear diffusively coupled systems for clarity, where unless otherwise stated, we use the combinatorial Laplacian matrix, $L=D-W$, with $D$ being the diagonal matrix of row sums of $W$.  A large class of such systems can be represented as
\begin{equation}
\label{system}
\dot{\vec{x}} = F(\vec{x})-\sigma  \left(L\otimes H\right) \vec{x}, 
\end{equation}
where $\vec{x}=\left[{\bf x}_1,{\bf x}_2,...,{\bf x}_N\right]$ denotes the set of state vectors for $N$ nodes, $F(\vec{x}) = \left[f({\bf x}_1), f({\bf x}_2), ..., f({\bf x}_N)\right]$ is the uncoupled dynamics, $\sigma$ is a global coupling strength, and $H$ represents a coupling function that can be used for multidimensional state vectors; again for simplicity, we will restrict our discussion here to the case of $H=\mathbb{I}_d$, something that is uncommon in the broader MSF literature, perhaps due to the simple linear form of the associated MSF (see Fig.~\ref{fig:MSF}). 

We note that coupling through the random walk or normalized symmetric Laplacian can be presented in a similar form by substitution of $L$, e.g., using $L_{rw}=\mathbb{I}_3-D^{-1}W$ or $\mathcal{L}=D^{1/2}L_{rw}D^{-1/2}$, however, since the present work is restricted to undirected networks, we will only consider the normalized symmetric Laplacian as an alternative. Importantly, the local scaling and overall bounded nature of the eigenvalues of $\mathcal{L}$ often lead to very different results for MSF stability analysis as we will see.

The vectorized equation~(\ref{system}) with $H$ being the identity then encodes $N$ coupled differential equations of the form
\begin{equation}
\label{single}
\dot{{\bf x}_i} = f({\bf x}_i)+\sigma \sum_{j}W_{ij} \left( {\bf x}_j-{\bf x}_i\right), 
\end{equation}
where the global coupling strength, $\sigma$, allows for varying the impact of coupling, in proportion to $D$, the node degrees, on the uncoupled dynamics. The main difference in using $\mathcal{L}$ being the uniform proportionality that is imposed by $\sigma$ on each node dynamics.

\subsection{Master Stability Function}
\label{sec:MSF}
The MSF approach to stability analysis proceeds by considering the eigen-decomposition of Equ.~(\ref{system}), 
under a subsequent change of coordinates into the eigenbasis of $L$ resulting in the $N$ uncoupled variational equations 
\begin{equation}
\label{variation}
\dot{\zeta}_i = \left[Df - KH\right]\zeta_i,
\end{equation}
which now represent the variation in the direction of each eigenvector of $L$ where $K$ plays the role of a $\sigma$-boosted eigenvalue, i.e. $K=\sigma\lambda$.  Assuming the coupled system is path connected, there is a single direction associated with the eigenvalue $\lambda_1=0$ of $L$, where $\left\{\lambda_i\right\}_{i=1}^N$ are the ordered eigenvalues of the combinatorial Laplacian associated with the coupling adjacency matrix; this direction is associated with the synchronization manifold. 

The MSF, often denoted as $\Psi(K)$, is then defined as the Largest Lyapunov Exponent (LLE) of the generic variational equation~(\ref{variation}) as a function of the parameter $K$.  In practice, $\Psi(K)$ is estimated numerically for a range of parameter values using a QR factorization~\cite{vonbremen1997efficient}, and the regions for which $\Psi(K)<0$ are estimated. See~\cite{huang2009generic} for a detailed analysis for many common chaotic oscillator dynamics under various coupling choices, where the systems are classified into types associated with the number of roots of $\Psi(K)$. Given the MSF and a network, if a single $\sigma$ value can be chosen such that $\Psi(\sigma\lambda_k)<0$ for all $k\in 2,\dots,N$, then the synchronization manifold for those dynamics coupled over that network is deemed linearly stable. 

\begin{figure}[ht!]
\begin{tikzpicture}[scale=0.95]
\node at (0,0) {\includegraphics[width=5.5cm]{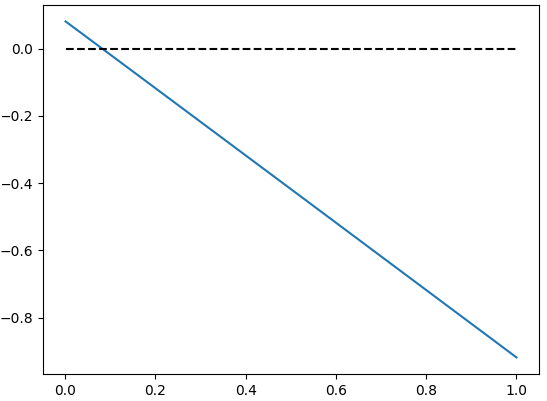}};
\node at (0.2,-2.5) {$K (\sigma\lambda)$};
\node[rotate=90] at (-3.25,0) {$\Psi(K)$};
\node at (0.2,2.75) {\small Master Stability Function};
\node at (0.2,2.35) {\small R\"{o}ssler systems with $H=\mathbb{I}_3$};
\end{tikzpicture}
\caption{The MSF for identical R\"{o}ssler systems ($a=b=0.2$ and $c=5.7$) under direct one-to-one variable coupling, i.e. $H=\mathbb{I}_3$ is a negative sloped line crossing at approximately $K_\alpha\approx 0.08$.\label{fig:MSF}}
\end{figure}

As an example, we consider identical R\"{o}ssler attractors (with the common parameter choices of $a=b=0.2$ and $c=5.7$) under full one-to-one variable coupling, i.e. $H=\mathbb{I}_3$.  This choice results in an MSF that is essentially a straight line having negative slope and crossing the $K$ axis once at $K_\alpha\approx 0.08$; the approximation is shown in Fig.~\ref{fig:MSF}.  The linearity of the MSF is likely related to the complete state information being shared under our choice of $H$, which means increased coupling generally leads to faster synchronization.  Again, other coupling functions may prove to be more interesting case studies and may lead to more complex bottleneck measures, but we leave their consideration for future work.

This particular MSF qualifies as Type I, wherein $\Psi(K)<0$ on $(K_\alpha,\infty)$ for some $K_\alpha>0$~\cite{huang2009generic}.  In this case (and the case of zero-entropy oscillators where $K_\alpha=0$), any connected network will be able to synchronize given a large enough global coupling strength.  Here, $\sigma$ can be seen as a dial on the rate of information exchange and dissipation in proportion to the network.  Not all Type I MSF have such simple form, but for such cases, in particular, the value of $K_\alpha$ does enable an approximation for the minimum coupling required for a linearly stable synchronization manifold, e.g., $\sigma*>0.08/\lambda_2$.  This is theoretically true in general for all Type I MSF, but the transient dynamics and overall rates of convergence may vary across the network leading to less straightforward analysis by information theory measures, which is why we confine our initial consideration to $H=\mathbb{I}_d$.

\subsection{The Cheeger Inequalities and Fiedler Vectors}
\label{sec:Iso} 
We now define both Cheeger constants after providing relevant notation used in~\cite{Chung96}.  Given a symmetric weighted graph $G=(V,E,W)$, we define the \textit{edge boundary} of a set $S\subset V$ to be 
\begin{equation*}
\label{partialS}
\partial S = \left\{\left\{u,v\right\}\in E : u\in S, v\notin S\right\},
\end{equation*}
and its size, $|\partial S|$, to be the sum of the edge weights, i.e. 
\begin{equation*}
|\partial S|=\sum_{\left\{u,v\right\}\in\partial S}{W_{u,v}}.  
\end{equation*}

\subsubsection{Isoperimetric Number}
The original isoperimetric number for graphs was presented as a \textit{Cheeger constant} in~\cite{mohar1989isoperimetric} where the basic relationship to the smallest non-zero eigenvalue of the combinatorial Laplacian matrix (the Cheeger inequality) was proven.  For historical reasons, there is not widespread agreement on notation; the most common being $h_G$, which was the notation inherited from Cheeger's work, but it has been used for both measures.  To prevent confusion, we will refer to this particular constant as the \textit{isoperimetric number}, and denote it explicitly as $\textrm{iso}_G$ as in~\cite{Nica18}; we define it as 
\begin{equation}
\label{hG}
\textrm{iso}_G = \min_{S\subseteq V}\frac{|\partial S|}{\min{(|S|,|\overline{S}|)}},
\end{equation}
where we note that $|\cdot|$ represents different operations depending on whether the argument is a set of vertices or edges.  

Solving for this graph property is NP-hard, but for a connected graph, its value can be bounded using the value of $\lambda_2$, i.e. the first non-zero eigenvalue of the combinatorial Laplacian matrix as 
\begin{equation}
\lambda_2/2\leq \textrm{iso}_G \leq\sqrt{2\Delta\lambda_2},
\end{equation}
where $\Delta$ is the maximum node degree in $G$~\cite{mohar1989isoperimetric}. 

Although the analogy to Cheeger's work holds in the limit of large graphs, for finite graphs, the combinatorial graph Laplacian matrix is more accurately understood as a Fokker-Plank operator rather than a pure Laplace-Beltrami operator, where the degrees of the nodes represent a drift term~\cite{nadler2005diffusion,coifman2006diffusion}.  This discrepancy led to the eventual definition of an alternative Cheeger constant in~\cite{Chung96} based on volumetric measures of subsets, which has since also been termed \textit{graph conductance} due to its relation to quantifying limits on flows.

\subsubsection{Graph Conductance}
The conductance of a graph, which we will denote by $\Phi_G$, uses the edge weights in the denominator of the defining ratio instead of the subset order to more directly relate this quantity with flows.  The resulting local degree scaling makes this matrix better align with the Laplace-Beltrami operator on finite graphs, which makes this graph property more central to modern spectral graph theory; though importantly, it is not related to the common graph Laplacian that is more often used in synchronization dynamics. 

The \textit{conductance} of $G$ is defined as 
\begin{equation}
\Phi_G = \min_{S\subseteq V}\frac{|\partial S|}{\min{(vol(S),vol(\overline{S}))}},
\end{equation}
where the volume of a subset $S\subset V$, i.e. $vol(S)$, is the sum of the degrees of vertices in $S$.  Similar to the isoperimetric number, the graph conductance is related to the eigenvalues of the random walk Laplacian matrix (and by algebraic similarity the normalized symmetric Laplacian matrix).  If we denote the eigenvalues shared by $L_{rw}$ and $\mathcal{L}$ as $\left\{\mu_i\right\}_{i=1}^N$, we then have a Cheeger inequality
\begin{equation}
\mu_2/2\leq \Phi_G \leq\sqrt{2\mu_2},
\end{equation}
where the main difference is the absence of $\Delta$, which is indicative of the normalization of the eigenvalues within the interval $[0,2]$. 

It is worth mentioning here that there are also so-called ``local'' Cheeger constants that seek to quantify the expansion (contraction) from (to) a subset of vertices within a larger graph, but they are just a restricted form of the Cheeger constants described above. The details of how these two bottleneck measures are related to their analog in Riemannian geometry is very interesting, but is well beyond the scope of this work; we refer the interested reader to~\cite{luxburg2004limits,hein2005graphs,nadler2005diffusion,coifman2006diffusion} for more details. 

\subsubsection{The Fiedler Vector}
The eigenvector associated with the smallest non-zero eigenvalue of the combinatorial Laplacian has become known as the Fiedler vector~\cite{fiedler1973algebraic}, but this term has similarly been used to describe the analogous vector for the normalized Laplacian.  In either instance, it is commonly used to estimate the graph cut that defines the associated Cheeger constant as it tends to change sign over that cut.  However, this approach often fails when there is hierarchical clustering or other ambiguity including small-set bottlenecks~\cite{clauset2004finding}.  In addition, if the algebraic connectivity has multiplicity, or even near-multiplicity, a higher dimensional Fiedler subspace will characterize this balanced cut, and the individual vector may only describe a subset of the important cut links. There are decentralized methods for approximating this vector~\cite{bertrand2013distributed}, but just as with the conductance, it better identifies the bottleneck to bulk synchronization.

\section{A Special Case: Bounding the Algebraic Connectivity by the Interlacing Theorem}
\label{sec:Inter}
As we move toward our definition, we will often rely on motivating examples where a pendant vertex is attached by a parameterized weighted edge to a larger connected weighted symmetric graph $G=(V,E,W)$, whose previous Laplacian spectra we denote as $\left\{\lambda_i\right\}_{i=1}^N$. This specific case provides a controlled experimental approach to the analysis of extreme small-set bottlenecks and an exploration of tipping points toward more balanced cuts, i.e. related to conductance. 

Given a graph $G$ as above, define a new graph $\tilde{G}$ by appending a single isolated node, $u$, to the vertex set $V$; the Laplacian spectrum of $\tilde{G}$ is then trivially $\left\{0\right\}\cup\left\{\lambda_i\right\}_{i=1}^N$.  Now, adding a weighted edge $\left\{u,v,\alpha\right\}$ connecting this isolated node ($u$) to some other node $v\in V$, we call this new graph $H$, and note that it is connected having order $|H|=N+1$. 

By the interlacing theorem from~\cite{grone1990laplacian} (THM 4.1), we have that the Laplacian spectrum of $H$, which we will denote by $\left\{\mu_j\right\}_{j=1}^{N+1}$, will satisfy $\mu_1=0$ and 
\begin{equation*}
    \lambda_{j-1}\leq\mu_j\leq\lambda_{j} \quad \textrm{for all}  \quad j\in\left\{2,\dots, N\right\},
\end{equation*}  
where we point out the shift in index from the classic result due to the association with the original graph $G$; this implies also that $\mu_{N+1}\geq\lambda_N$.  From this, we know the algebraic connectivity of $H$ (i.e. $\mu_2$) both controls the stability of the synchronization manifold for dynamics having a Type I MSF and is bounded above by $\lambda_2$, the algebraic connectivity of the original graph $G$. Further, for small $\alpha>0$, it can be shown that the growth of $\mu_2$ is bounded from above by $\frac{N+1}{N} \alpha$.  

To fully understand the relevant lower bound, we must consider the attachment point, $v\in V$, and its wider role in the diffusion process on $G$.  Consider the resolvent Green's function for the original graph Laplacian, given by $G(\lambda)=(L-\lambda \mathbb{I})^{-1}$, having domain $\lambda\in\mathbb{C}\setminus \left\{\lambda_i\right\}_{i=1}^N$.  The diagonal elements of this matrix are known to encode the responses to perturbations at each node in the graph $G$.  If we expand this function about the trivial eigenvalue $\lambda_1=0$, we obtain an approximation for the diagonal elements: 
\begin{equation}
G(\lambda)_{vv} = \frac{-1}{N \lambda} + r_v +\mathcal{O}(\lambda),     
\label{equ:greens}
\end{equation}
where $r_v = \sum_{k=2}^N\frac{|\varphi_k(v)|^2}{\lambda_k}$ with $\varphi_k$ being the Laplacian eigenvector associated with $\lambda_k$.  These static contributions to the sum are exactly the diagonal elements of the Moore-Penrose pseudoinverse, and they represent the overall response of that node to global perturbations~\cite{fouss2007random} (which is why the effective resistance matrix was referenced in~\cite{dorfler2010synchronization}). 

From low-rank perturbation theory the eigenvalue problem for $L(H)=L_H$ can be written as a function of the edge weight, $\alpha$, as:
\begin{equation*}
L_H(\alpha)\vec{x} = \left(\begin{array}{c|c}
    L_G + \alpha e_v e_v^T & -\alpha e_v \\
 \hline 
 -\alpha e_v & \alpha 
\end{array}\right) \cdot \left(\begin{array}{c}
    y\\
 \hline z 
  \end{array}\right)
=\lambda \left(\begin{array}{c}
    y\\
  \hline  z 
\end{array}\right),
\end{equation*}
where $e_v$ is the indicator vector of node $v$ and $y$ is a vector of length $N$.  Solving the bottom equation for $z$ and substituting into the larger modified eigenvalue equation, we get $z=\frac{\alpha}{\alpha-\lambda} y_v$ and $(L_G+\alpha e_v e_v^T)y - \alpha e_v z = \lambda y$, which combine and simplify as:
\begin{equation*}
\begin{aligned}    
(L_G-\lambda\mathbb{I}) y &= \frac{\alpha^2}{\alpha-\lambda} e_v y_v - \alpha e_v e_v^Ty \\
    (L_G-\lambda\mathbb{I}) y &= \frac{\alpha \lambda}{\alpha-\lambda} e_v y_v. \\  
    \end{aligned}
\end{equation*}
Multiplying through by the resolvent $(L_G-\lambda\mathbb{I})^{-1}$ and considering the $v$-th component, we obtain the secular equation that describes the effect of the perturbation on node $v$ as
\begin{equation}
\label{equ:sec}
G(\lambda)_{vv}= \frac{\alpha-\lambda} {\alpha \lambda} . 
\end{equation}
Substituting the approximated right hand side of Equ.~(\ref{equ:greens}) into~(\ref{equ:sec}) and solving for $\lambda$, we find
\begin{equation*}
\lambda \approx \frac{N+1}{N}\frac{\alpha}{1+\alpha r_v}.
\end{equation*}
Now considering the loose bound 
\begin{equation*}
r_v = \sum_{k=2}^N \frac{|\varphi_k(v)|^2}{\lambda_k}\leq \frac{1}{\lambda_2}\sum_{k=2}^N|\varphi_k(v)|^2 < \frac{1}{\lambda_2},
\end{equation*}
we can define both an upper and lower bound for the algebraic connectivity, $\mu_2$, of the new graph $H$ (valid for small values of $\alpha$) as
\begin{equation}
\frac{N+1}{N}\frac{\alpha}{1+\alpha/\lambda_2}\leq \mu_2\leq \min{\left\{\lambda_2,\frac{N+1}{N}\alpha\right\}}.
\end{equation}

This result from spectral graph theory provides an excellent guide for assessing our bottleneck measure as it should closely follow this same saturation for the special case as the synchronization bottleneck transitions from being the pendant edge to a broader cut that may be more closely associated with the conductance of the graph $G$, but perturbed by the additional volume in $H$. 

There is a natural scale of the so-called secular equation, Equ.~(\ref{equ:sec}), which is given by $\alpha* \sim 1/r_v$.  When $\alpha<<\alpha*$, the estimate is solidly in the linear regime, i.e. $\mu_2\sim \frac{N+1}{N}\alpha$, and the pendant bottleneck dominates; but, once $\alpha>>\alpha*$, the connectivity will saturate as $\mu_2\sim \frac{N+1}{N r_v}$, which can be estimated in terms of the algebraic connectivity of the original graph $G$. Here, the pendant bottleneck may or may not play a role as the bottleneck becomes dominated by the structure of $G$.  

The shift between limiting bottlenecks is often gradual as $\alpha$ increases and is highly dependent on the attachment vertex $v\in V$.  So, although our measure will identify a single minimizer cut, it is not always that cut that is the true limitation to complete synchrony. 

Despite this ambiguity, the special case of pendant vertex attachment provides a reasonable means of testing how well the proposed bottleneck measure captures the true transition, and under what circumstances we expect it to fail. It must remain clear that the true transition point of the limiting bottleneck can be a global network feature, and thus does not lend itself to being defined by a simple local information optimization problem. Despite this limitation, we feel it is valuable to define a useful approximation based entirely on local information that can provide insight into decentralized system optimization.

\section{The Synchronization Bottleneck Measure}
\label{sec:Xi}
We require one additional concept for our definition; the (outside) \textit{vertex boundary} of a set $S\subset V$, which we define as 
\begin{equation}
\label{deltaS}
\delta S = \left\{v\notin S : \left\{u,v\right\}\in E,u\in S\right\}.
\end{equation}

\begin{figure}[ht!]
\begin{tikzpicture}
\node at (0,0) {\includegraphics[width=7.5cm]{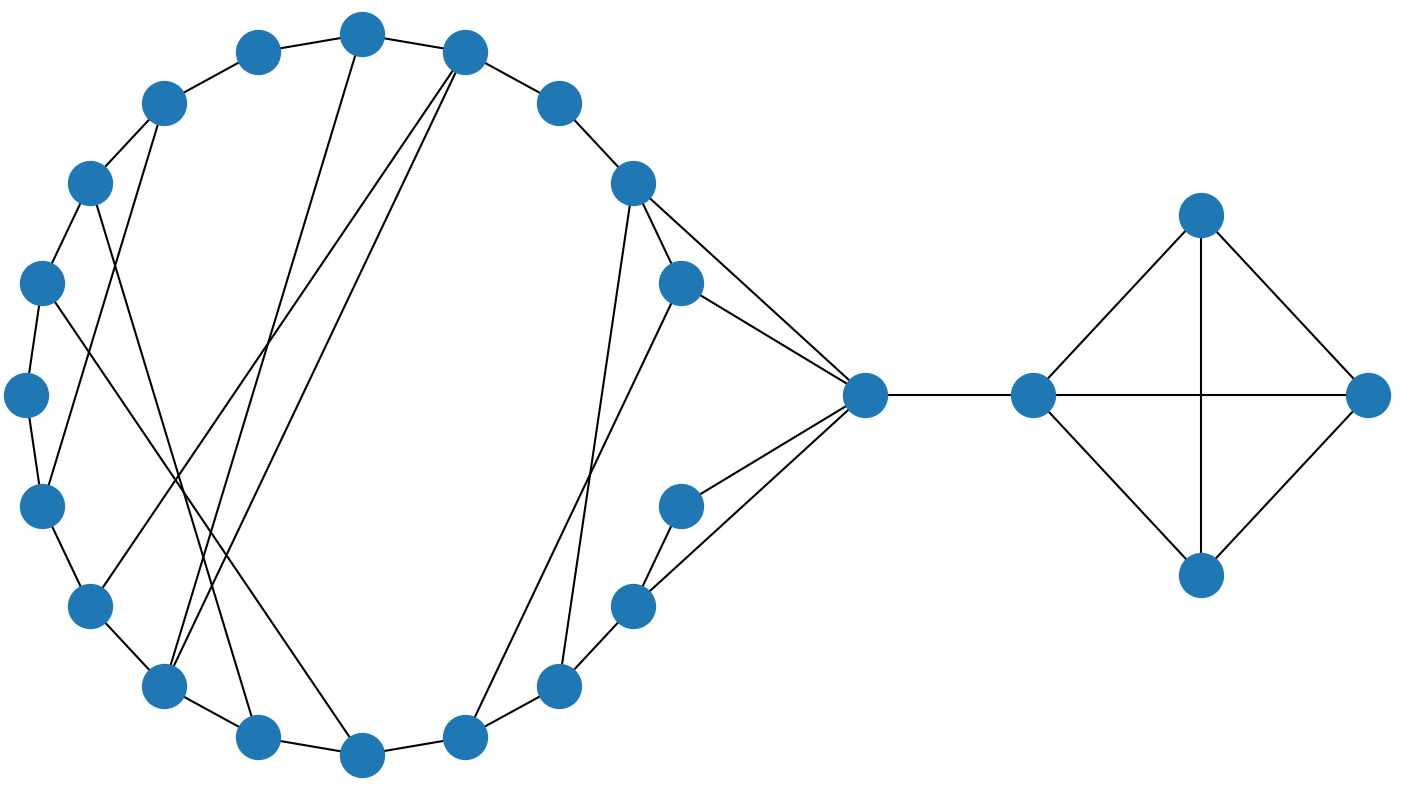}};
\draw[black,fill=orange, opacity=0.4] (0.9,0) ellipse (0.2cm and 0.5cm) node (A) {};
\draw[green,fill=green, opacity=0.4] (1.35,0) ellipse (0.1cm and 0.75cm) node (B) {};
\draw[dashed] (1.35,-0.5)--(1.35,.5);
\draw[black,fill=magenta, opacity=0.4] (1.8,0) ellipse (0.2cm and 0.5cm) node (C) {};
\draw[orange,fill=orange, opacity=0.4] (1.05,2.25) rectangle (1.85,2.5);
\node at (2.15,2.375) {$\delta \overline{S}$};
\draw[green,fill=green, opacity=0.4] (1.05,1.75) rectangle (1.85,2);
\node at (2.15,1.875) {$|\partial S|$};
\draw[magenta,fill=magenta, opacity=0.4] (1.05,1.25) rectangle (1.85,1.5);
\node at (2.15,1.375) {$\delta S$};
\draw (1,1.15) rectangle (2.5,2.6);
\draw[gray,fill=gray, opacity=0.2] (-1.4,0) ellipse (2.5cm and 2.2cm) node (A) {};
\node at (0,-1.3) {\Large $S$};
\end{tikzpicture}
\caption{A simple graph $G=(V,E)$ constructed from a Small-World (SW) graph (with high rewiring parameter, but low conductance) and a complete graph on four vertices appended; a particular subset $S\subset V$, the edge boundary of $S$, i.e. $\partial S$, and the outer vertex boundaries of $S$ and $\overline{S}$ are indicated in gray, green, red, and orange, respectively.\label{fig:defns}}
\end{figure}

Figure~\ref{fig:defns} provides a simple (unweighted) graph having an unbalanced bottleneck between the set $S$ (highlighted in gray) and its compliment $\overline{S}$ to illustrate the relevant quantities.

The \textit{synchronization bottleneck ratio} is defined for any bond of a connected symmetric graph $G=(V,E,W)$, i.e. a cut set $C=(S,\overline{S})$ that results in exactly two connected components, as
\begin{equation}
\Xi(G,C)=\Xi(G,(S,\overline{S}))= \frac{2|\partial S|}{vol(\delta S)+vol(\delta \overline{S})}.
\end{equation}
Note that this is simply the weighted size of the cut divided by the average between the volumes of boundary nodes on either side of that cut. Since $vol(\delta \overline{S})$ and $vol(\delta S)$ will each include the weight of the cut $C=(S,\overline{S})$, i.e. $|\partial S|$, by definition, the ratio satisfies $0<\Xi(G,C)\leq 1$. We also note the symmetry in this measure, indicating that we need only consider subsets of vertices up to one half of the order of the graph to obtain all relevant cuts.

Letting $B^*(G)\subset C^*(G)$ be the subset of the cut set that includes all bonds of the graph $G$, i.e. the set of all graph cuts that result in two connected components. We then define the \textit{synchronization bottleneck measure} by minimizing this ratio over that set of bond cuts,
\begin{equation}
\begin{aligned}
\Xi_G&=\min_{C\in B^*(G)}{\Xi(G,C)}\\
&=\min_{(S,\overline{S})\in B^*(G)}\frac{2|\partial S|}{vol(\delta S)+vol(\delta \overline{S})}.\\
\end{aligned}
\end{equation}

\section{Validating the Measure}
\label{sec:results}
\label{sec:validate}
Although simply stated, we can first validate this definition by considering the synchronization dynamics in Equ.~(\ref{system}) for a simple path graph on two vertices under a global coupling strength of $\sigma$, resulting in the overall scaled coupling network shown in Fig.~\ref{fig:P2}.

\begin{figure}[ht!]
\centering
\begin{tikzpicture}[scale=1]
\draw[black] (2,0) circle (10pt) node (C) {};
\draw[black] (4,0) circle (10pt) node (E) {};
\draw[black] (3,-0.5) node {$\sigma$};
\draw[line width = 0.25mm, double distance = 0.5mm, line cap=rect] (C) -- (E);
\draw[black, fill=white] (2,0) circle (9pt) node (C) {$A$};
\draw[black, fill=white] (4,0) circle (9pt) node (E) {$B$};
\end{tikzpicture}
\caption{A simple path graph on two vertices with a global coupling strength of $\sigma$ results in a coupling network with only one cut having $|\partial S|=\sigma$.\label{fig:P2}}
\end{figure}
Obviously, here we see that $\Xi_G = 2\sigma/(\sigma+\sigma)=1$ for all choices of $\sigma$, fully characterizing one extreme value of the measure (i.e. no bottleneck at all). The other extreme is disallowed by our measure under the assumption of connectedness.  We now consider several graphs that were chosen to further validate specific features of our definition.

\subsection{Distinction from Cheeger Constants}
Consider the graph from Fig.~\ref{fig:defns}, reproduced here with node labels, which was designed to illustrate that the synchronization bottleneck is not necessarily defined by either Cheeger constant cut. We combined a nearest neighbor Small-World (SW) graph having $N_1=20$ nodes with a complete graph on $N_2=4$ vertices by connecting them via a single edge, resulting in a graph with order $|G|=N_1+N_2=24$. Table~\ref{table:defns} displays the values for $\Xi_G, \textrm{iso}_G,$ and $\Phi_G$ for this graph alongside the node sets that define each of these measures.  Note that all measures are symmetric, meaning the values can be computed by a brute force search of subsets of nodes up to size $\lceil N/2\rceil$ to ensure accuracy.

\begin{figure}[ht!]
\includegraphics[width=8cm]{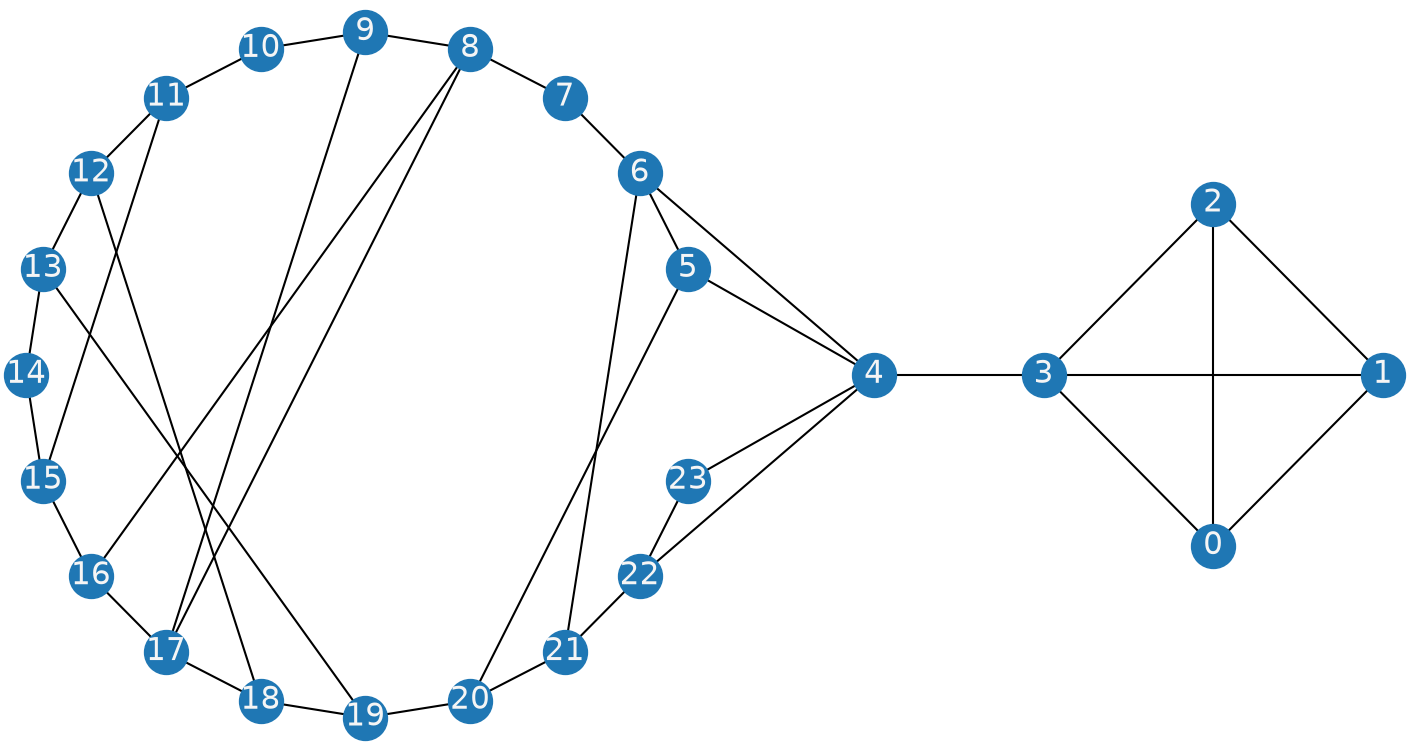}
\end{figure}
\begin{table}[ht!]
\caption{Bottleneck measures for the graph in Fig.~\ref{fig:defns} along with the cuts that define them.\label{table:defns}}
\renewcommand{\arraystretch}{1.5} 
\begin{tabular}{|c|c|}
\hline
\textbf{Graph Property}&\textbf{Ratio Minimizer Set} $S$\\ 
\hline\hline
$\Xi_G=2/9$&$\left\{0, 1, 2, 3\right\}$ \\
\hline
$\textrm{iso}_G=1/6$ & $\left\{0, 1, 2, 3, 4, 5, 6, 7, 20, 21, 22, 23\right\}$\\
\hline
$\Phi_G=1/18$&$\left\{0, 1, 2, 3, 4, 5, 6, 20, 21, 22, 23\right\}$\\
\hline
\end{tabular}
\end{table}

Here, we simply note that $\Xi_G$ is in fact defined by the cut that separates the complete graph from the rest of the vertices.  Additionally, we see that the two Cheeger constants are, in this case, defined by different sets entirely, where the optimal subsets had either a larger order or volume as the case may be.

\subsection{The Tipping Point in a Weighted Path Graph}
Since the synchronization bottleneck measure is entirely defined in terms of edge weights, a parameterized weighted path graph on five vertices suffices to explore a large set of interesting cases (including the previous graph as a form of reduced order model).  Figure~\ref{P5} shows an example where one of the links maintains a weight of unity and plays the role of the edge cut that defines the graph conductance $\Phi_G$ by design; while the three other links take on parameterized values of $w_\ell, w_c$ and $\alpha$ (subscripts of $w$ indicating left and center, respectively) where we assume $w_\ell>w_c>>1>\alpha$ in order to ensure that the two cuts that define $\Xi_G$ and $\Phi_G$ are not equal. 
\begin{figure}[ht!]
    \centering
\begin{tikzpicture}[scale = 0.9]
\draw[black] (0,0) circle (10pt) node (A) {};
\draw[black] (2,0) circle (10pt) node (B) {};
\draw[black] (4,0) circle (10pt) node (C) {};
\draw[black] (6,0) circle (10pt) node (D) {};
\draw[black] (8,0) circle (10pt) node (E) {};
\draw[black] (1,-0.5) node {$w_\ell$};
\draw[black] (3,-0.5) node {$1$};
\draw[black] (5,-0.5) node {$w_c$};
\draw[black] (7,-0.5) node {$\alpha$};
\draw[line width = 0.25mm, double distance = 1.25mm, line cap=rect] (B) -- (C);
\draw[line width = 0.25mm, double distance = 4mm, line cap=rect] (A) -- (B);
\draw[line width = 0.25mm, double distance = 3mm, line cap=rect] (C) -- (D);
\draw[line width = 0.25mm, double distance = 0.5mm, line cap=rect] (D) -- (E);
\draw[black, fill=white] (0,0) circle (9pt) node (A) {$A$};
\draw[black, fill=white] (2,0) circle (9pt) node (B) {$B$};
\draw[black, fill=white] (4,0) circle (9pt) node (C) {$C$};
\draw[black, fill=white] (6,0) circle (9pt) node (D) {$D$};
\draw[black, fill=white] (8,0) circle (9pt) node (E) {$E$};
\end{tikzpicture}
    \caption{A weighted path graph on five vertices, which can represent the bottlenecks for dynamics coupled on many graphs when $w_\ell>w_c>>1>\alpha$.\label{P5}}
\end{figure}

If we further assume $ w_\ell>w_c+\alpha$ and $\alpha>1/2$, then it is straight forward to show that $\textrm{iso}_G=1/2$ and $\Phi_G = 1/(1+2w_c+2\alpha)$, both relying on the minimizer set $S=\left\{A,B\right\}$ with the minimum cut being the single edge $\left\{B,C\right\}$. It may be instructive to note that if $\alpha<1/2$, $\textrm{iso}_G$ would then be defined in terms of the cut $\left\{D,E\right\}$, while $\Phi_G$ would remain defined by the cut $\left\{B,C\right\}$.  In addition, we note that while $\textrm{iso}_G$ would then capture a change in the bottleneck for complete synchronization in this case, it is defined in terms of the number of nodes, and thus its transition point with the parameter $\alpha$ has no relation to either of the values $w_\ell$ or $w_c$, meaning it is of no further interest for flow analysis. 

This example can be viewed as the attachment of a parameterized pendant vertex by holding $w_\ell=4$ and $w_c=2$ fixed and allowing $\alpha$ to vary.  If we now compute the synchronization bottleneck ratio, $\Xi(G,C)$, for each of the possible graph cuts $C\in B^*(G)$, we obtain the expressions given in Table~\ref{xi}; example values are provided for the specific choice of $\alpha=1/4$ in the final column.  For this chosen value of $\alpha=1/4$, we have the minimizer set $S=\left\{A,B,C,D\right\}$ --- or by symmetry $S=\left\{E\right\}$ --- and the graph cut defining the bottleneck being the single edge $\left\{D,E\right\}$ having weight $\alpha$. 

\begin{table}[hbtp!]
\caption{The computation of the synchronization bottleneck ratio, i.e. $\Xi(G,(S,\overline{S}))$, for all graph cuts $C=(S,\overline{S})$ of the weighted path graph where $w_\ell>w_C+\alpha$ and $\alpha<1$, along with an Example value for the case where $w_\ell=6, w_c=4$ and $\alpha=1/2$ for which $\Xi_G$ and $\Phi_G$ are defined by different cuts.\label{xi}}
\renewcommand{\arraystretch}{1.75}
\scriptsize
\begin{tabular}{|c|c|c|c|c|c|c|}
\hline
$S$ & $\partial S$& $|\partial S|$ &$\delta \overline{S}$  & $\delta S$ &$ \Xi(G,(S,\overline{S}))$&Ex: $\alpha=1/4$\\
\hline
\scriptsize$\left\{A\right\}$ & \scriptsize$\left\{A,B\right\}$ & $w_\ell$ & $\left\{A\right\}$& $\left\{B\right\}$ & $\frac{2w_\ell}{2w_\ell+1}$& 8/9\\
\scriptsize$\left\{A,B\right\}$ & \scriptsize$\left\{B,C\right\}$& $1$ & $\left\{B\right\}$& $\left\{C\right\}$& $\frac{2}{w_\ell+w_c+2}$ & 1/4 \\
\scriptsize$\left\{A,B,C\right\}$ & \scriptsize$\left\{C,D\right\}$& $w_c$ & $\left\{C\right\}$& $\left\{D\right\}$ & $\frac{2w_c}{2w_c+\alpha+1}$& 16/21\\
\scriptsize$\left\{A,B,C,D\right\}$ & \scriptsize$\left\{D,E\right\}$& $\alpha$ & $\left\{D\right\}$& $\left\{E\right\}$& $\frac{2\alpha}{2\alpha+w_c}$& \textbf{1/5} \\
\hline
\end{tabular}
\end{table}

\begin{figure}[ht!]
\includegraphics[width=8cm]{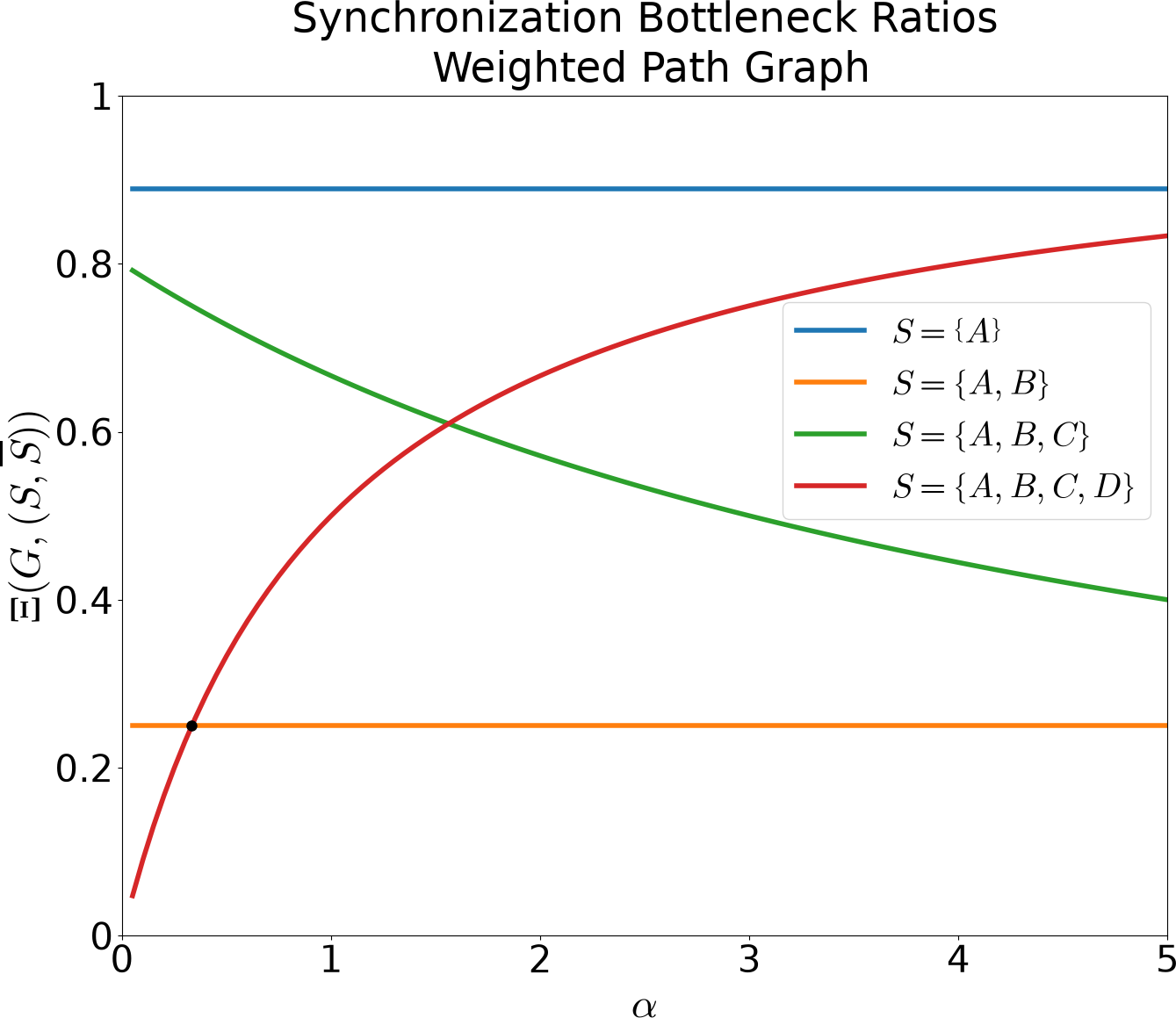}
\caption{The synchronization bottleneck ratio is plotted for all four possible cuts of the weighted path graph from Fig.~\ref{P5} with $w_\ell=4$ and $w_c=2$, while varying $\alpha$; the important intersection being located at $\alpha=1/3$ where the minimizing set changes from $\left\{A,B,C,D\right\}$ to $\left\{A,B\right\}$ resulting in a value of $\Xi_G=1/4$ for $\alpha>1/3$. \label{fig:Xi_paths}}
\end{figure}

Keeping $w_\ell=4$ and $w_c=2$ held constant, we now plot the synchronization bottleneck ratio for all four cuts of this graph as a function of the parameter $\alpha\in [0,5]$ in Fig.~\ref{fig:Xi_paths}.  Since our measure is defined by minimizing this argument, we find that the synchronization bottleneck measure reaches a transition point when $\alpha=1/3$.  

Thus, our measure predicts that the same bottleneck identified for the Example case of $\alpha=1/4$ will hold for all such values $\alpha<1/3$.  For values $\alpha>1/3$, we then expect the cut $\left\{B,C\right\}$ associated with the conductance to define the limiting synchronization bottleneck for this graph.  It should be pointed out also that for extreme values of $\alpha>>w_\ell$, the bottleneck will eventually be defined by the cut $\left\{C,D\right\}$ as the green curve falls below the orange line in the limit.  These kinds of transition points from one limiting set to another are generally observed as corner points in the subsequent plots of our measure, e.g. Fig.~\ref{fig:alpha}.

We reiterate that each of these curves represents a potential bottleneck, and as two curves approach an intersection, they both act as competing limiting bottlenecks. Since our measure is local, we do not claim that this transition point is the true inflection point, as that will depend on more global dynamic features including how much entropy is generated on each side of the graph cuts and the existing dissipation within each set.  However, under our assumption of high quality information coupling, i.e. $H=\mathbb{I}_3$, and the simple path structure here, we will find that this transition point is reliable in this case.

\begin{figure}
\includegraphics[width=7.5cm]{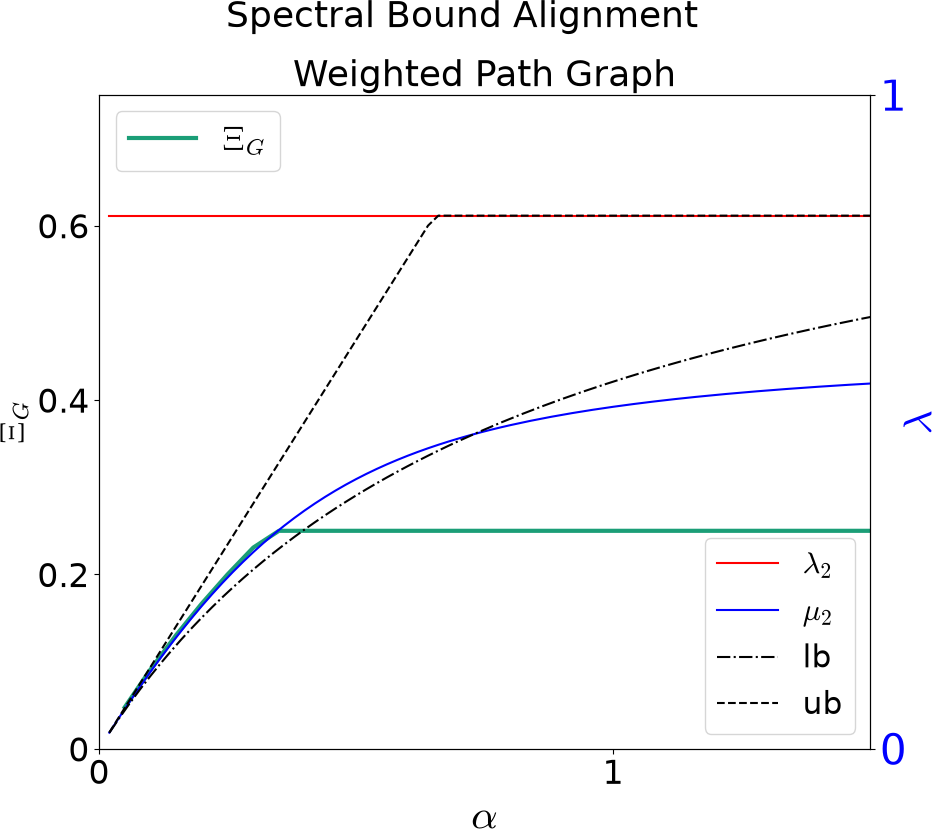}
\caption{The general shape of $\Xi_G$ as a function $\alpha$ indicates an close relationship with the algebraic connectivity, which is shown on the re-scaled axis along with its upper and lower bounds for small $\alpha$, which were obtained from the interlacing theorem; in this case, the predicted tipping point is relatively accurate.\label{fig:mu_2}}
\end{figure}

We can explore the concurrence of our measure with the change in limiting bottleneck from the pendant to the more balanced cut by considering the shape of $\Xi_G(\alpha)$ in comparison to the algebraic connectivity of graph as $\alpha$ varies along with the upper and lower bounds obtained in Sec.~\ref{sec:Inter} in Fig.~\ref{fig:mu_2}, where the scaling has been \textit{arbitrarily chosen for illustrative purposes}.  

In this particular case, we see a clear potential alignment with the algebraic connectivity of the graph for small $\alpha$. Although not necessarily accurate, our predicted transition point seems reasonable as a candidate parameter value where the true bottleneck begins to include the larger conductance-related cut. This provides a level of theoretical support for our measure as an approximation, and provides an indication for future work in bounding this measure by spectral means.

\subsection{Ramanujan graph with $\alpha$-weighted pendant}
\label{sec:alpha}
\begin{figure}[ht!]
\begin{tikzpicture}[scale=2]
\node at (0,0) {\includegraphics[width=4cm]{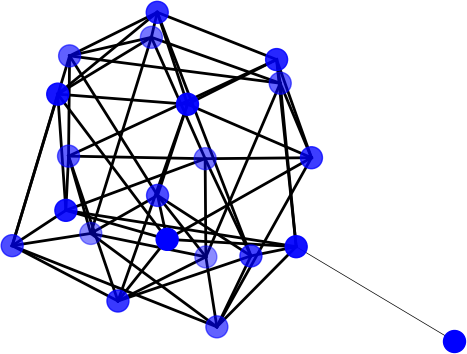}};
\node at (0.65,-0.35) {\large $\alpha$};
\draw[ultra thick, gray, draw opacity=0.25,line width=0.15cm] (0.25,-0.285) -- (0.95,-0.71);
\end{tikzpicture}\\
(a)\\
\begin{tabular}{cc}
\includegraphics[width=3.75cm]{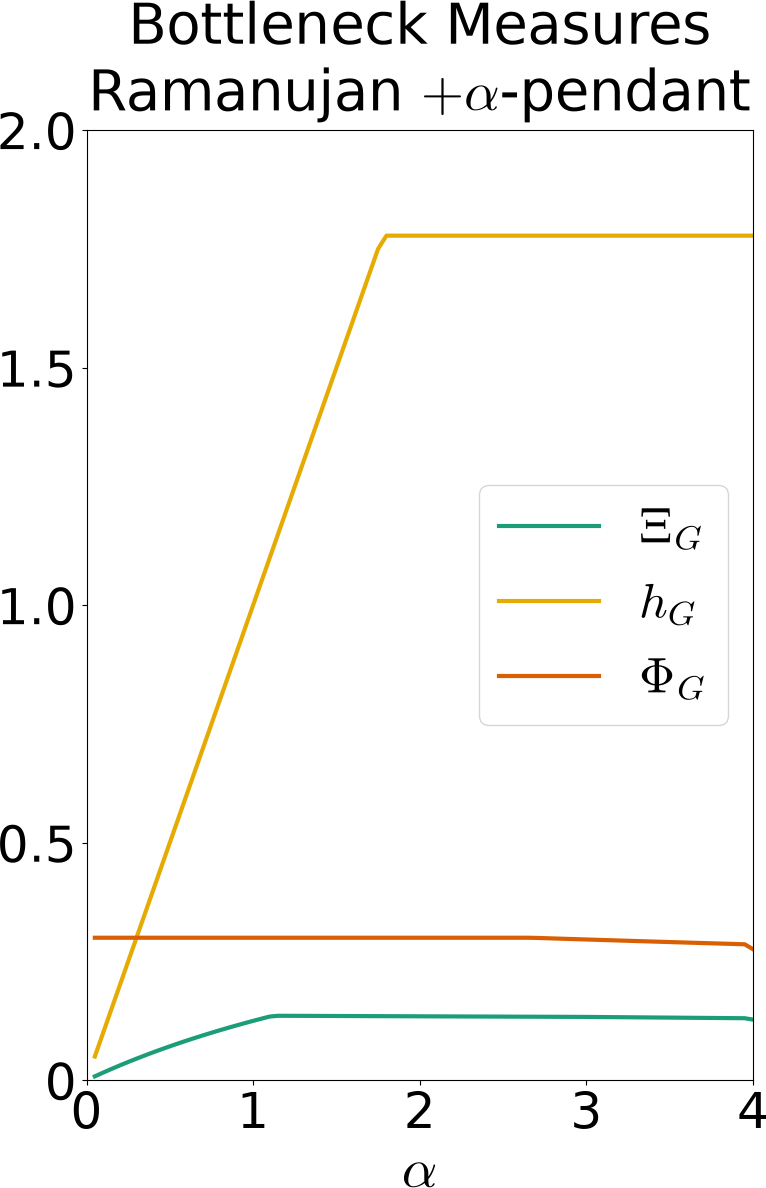}
 &
\includegraphics[width=4.25cm]{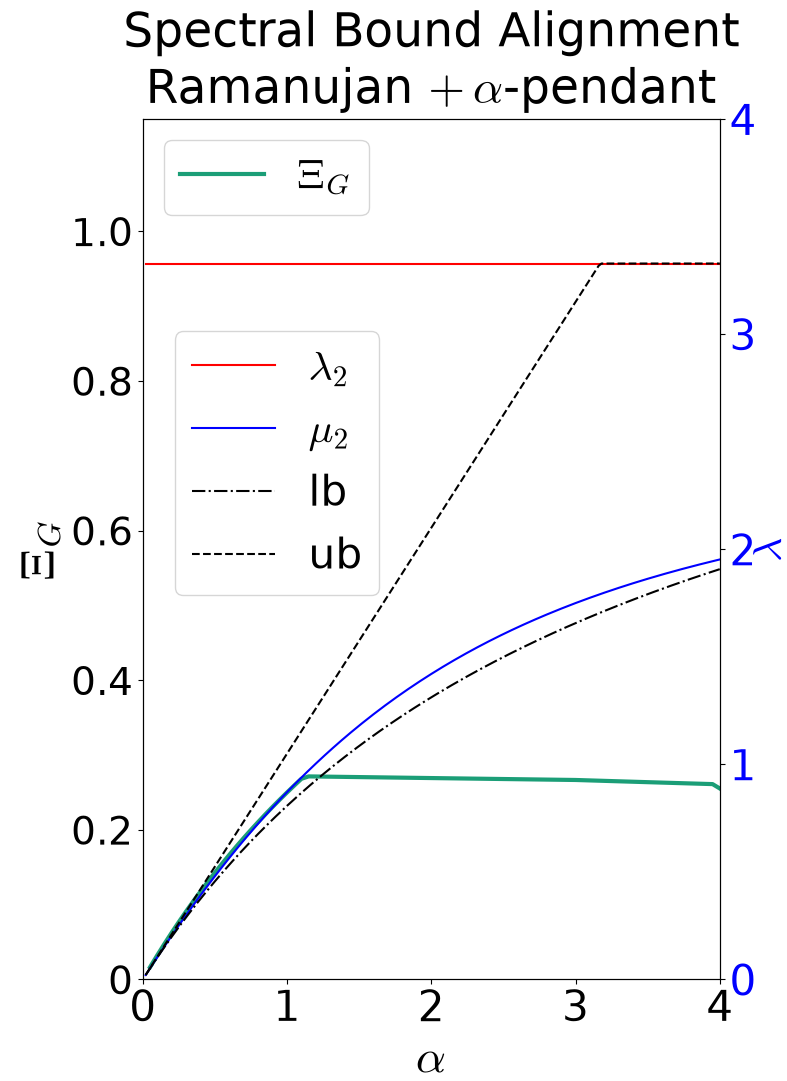}\\
(b)& (c)\\
\end{tabular}
\caption{(a) A simple parameterized model where a well connected $d$-regular graph has a parameterized weighted edge to an additional pendant vertex; allowing the parameter $\alpha$ to vary, we plot of the various bottleneck measures in (b) and compare the value of $\Xi_G$ with the algebraic connectivity in (c) along with the derived upper (ub) and lower (lb) bounds from Sec.~\ref{sec:Inter} by the interlacing theorem.\label{fig:alpha}}
\end{figure}

Now, consider attaching a pendant to a significantly different network structure, a Ramanujan expander graph where all of the original edges have weight one. Starting with a $d$-regular graph on $N=20$ nodes having a minimal bottleneck and $d=6$, we append a single pendant vertex to this graph with a parameterized edge weight $\alpha$ as shown in Fig.~\ref{fig:alpha} (a).  We then allow this parameter to vary over the range $a\in\left[0,4\right]$ and we plot the three bottleneck measures $\Xi_G, \textrm{iso}_G,$ and $\Phi_G$ in Figure~\ref{fig:alpha} (b). These are computed by brute force calculation, where the set $S$ that defines each of these measures may change as the parameter changes (noticeable as corner points in the plot). Note that due to the high connectivity of the underlying graph $G$, $\Xi_G(\alpha)$ remains low and transitions relatively early given the higher degree of all other nodes.

We also overlay a plot of $\Xi_G$ in (c) with a plot of the algebraic connectivity, $\mu_2$, again using a separate scale on the right that is chosen to align our measure with this curve.  We include the upper and lower bounds obtained from the interlacing theorem in Sec.~\ref{sec:Inter} to show the correspondence.

The first thing to note is that the graph conductance is relatively consistent, indicating why $\Phi_G$ is considered a robust measure of the main bottleneck.  In contrast, $\textrm{iso}_G$ is less reliable; although $\textrm{iso}_G$ indicates the synchronization bottleneck for smaller values of $\alpha$, there is no reason to believe that the parameter value where the minimizer changes has any meaning with respect to synchronization or information flow due to its definition in terms of order.  

For small $\alpha$, we see that $\Xi_G = \frac{2\alpha}{deg(v)+\alpha}$ where $v$ is the vertex to which our pendant is connected, and since $deg(v)$ includes $\alpha$, the slope will be less than or equal to one; here in fact, since each node has on average degree six, we can expect the initial values of $\Xi_G$ to remain below the line through the origin having slope $1/6$.  In general, $\Xi_G(\alpha)$ will grow sublinearly in $\alpha$ with its initial slope bounded by $1/deg(v)$ from the original graph. 

The lower bound for the new algebraic connectivity, $\mu_2$, derived in Sec.~\ref{sec:Inter}, also grows sublinearly with $\alpha$, however its slope is bounded from above by $(N+1)/N$ (with respect to the scale on the right axis).  Although we do not expect this measure to be exact, this functional relationship to the bounds on the new algebraic connectivity is a strong indicator for its validity, and offers a future potential connection to its own Cheeger-like inequality.  Further, the particular synchronization bottleneck ratio that is associated with the pendant cut will approach one, as $\alpha\rightarrow\infty$, meaning it will only remain the bottleneck if the original $G$ was an isolated node prior to attachment. 

Despite the concavity of the curve and its tendency toward saturation, we will generally encounter a sharp transition (corner point), where a new graph cut becomes the minimizer of the synchronization bottleneck ratio.  As before, we emphasize that while the transition for the minimizer may be abrupt and discrete, this will not necessarily be apparent from simulated experiments due to simulation and convergence rates.  The prior bottleneck does not simply disappear either. Each potential bottleneck in the graph becomes increasingly important to limiting flow as the synchronization bottleneck ratio associated with its cut reduces toward becoming the minimum.  

We note that here, the algebraic connectivity increases at a much slower rate related to the high dimensionality of the Fiedler subspace for the Ramanujan graph.  This means that the transition point for our measure occurs much earlier in the rise of $\mu_2$ in relation to the degree of the attachment vertex, and thus our predicted transition may not be as accurate.  In fact, due to the redundancy of the Fiedler subspace for the highly connected Ramanujan graph, the pendant cut will likely dominate as the observed bottleneck for much larger values of $\alpha$, though again it is hard to experimentally verify since the simulation time $T$ becomes an important factor.  If we are interested in the question of whether the system will return to the synchronous state ``rapidly'' then the predicted tipping point is relatively accurate still.  We now turn to numerical simulations to further validate the measure, where possible.

\section{Numerical Validation through Information Measures}
\label{sec:Disc}
\label{sec:verify}
Numerical simulations were carried out using identical R\"{o}ssler systems (with the common parameter choices of $a=b=0.2$ and $c=5.7$) under linear diffusive coupling on the networks above using the coupling function defined by $H=\mathbb{I}_3$. As noted in Subsec.~\ref{sec:MSF}, this particular choice of R\"{o}ssler systems and the one-to-one variable coupling scheme has a simple linear Type I MSF (see Fig.~\ref{fig:MSF}).  This means that in each simulation, a global coupling strength $\sigma$ can be chosen such that the dynamics exhibit a linearly stable synchronization manifold, i.e. $\sigma\lambda_2>K_\alpha\approx 0.08$.  However, we will mostly intentionally choose values for $\sigma$ such that this threshold is not met in order to explore the breakdown of synchronization due to insufficient information flow.

\subsection{Identifying Bottlenecks using Mutual Information Heatmaps}
The pairwise Mutual Information (MI) of networked node dynamics has been used as a measure of synchrony for decades~\cite{manrubia1999mutual,kreuz2007measuring,tirabassi2015inferring,bara2026redundant}.  Due to the variation in estimator biases and the particular details of how a continuous state space is discretized,  the actual numerical value of any given MI estimate is of little general interest; but, when used for pairwise comparison under the same parameter settings, MI provides a powerful indicator that is less skewed by extreme values and can incorporate important non-linear relationships~\cite{kraskov2004estimating,kreuz2007measuring}.  Therefore, for each system of networked dynamics, we provide heat maps, where high MI indicates stronger alignment of the underlying trajectories, and lower values indicate longer transient dynamics and/or a failure to synchronize altogether. 

Complete synchrony of the system should be observed through traditional MI estimates as a relatively consistent value across all pairs.  This historically desirable feature of information measures like MI is due to the tendency of traditional estimators to statistically de-emphasize outliers; in the case of synchronization dynamics, the short initial transient dynamics that may occur prior to convergence are considered the outliers in the context of the synchronization manifold.

However, this often celebrated feature of MI can actually lead to ambiguity when applied in higher dimensional settings, especially when applied to perturbations far from the manifold or under coupling strengths on the border of stability. There is nothing inherently problematic for using information measures in these contexts, but in order to observe the post transient behavior, longer trajectories are needed to allow for sufficient convergence.  Even in three dimensional state spaces, e.g., R\"{o}ssler systems, the computational complexity of geometric estimators place practical limits on the lengths of trajectories that are reasonable to compute, and so it has become common practice to simply limit the perturbation size and rely on relatively large coupling strengths so that more limited data can saturate at the synchronization manifold.

While this is sufficient to prove linear stability, it is not ideal for studying the path to synchrony, especially in low coupling strength regimes where nodes may fail to synchronize after complex transients.  Since we seek to intentionally choose coupling strengths for each example at or below the MSF-defined threshold, we expect significant pairwise differences in the node dynamics to persist.  Under these circumstances, we find that the popular Kraskov–St\"{o}gbauer–Grassberger (KSG) estimator, which is based on $k$-nearest neighbors will struggle to show the features of interest clearly without long trajectories, but the computational complexity then becomes a challenge. Instead, our analysis will focus on a new form of measure-based mutual information we call Geometric Partition Mutual Information (GPMI), which was introduced by the authors in~\cite{diggans2025generalizing}.  Since this new approach is still novel, we will begin with a comparison of results with the KSG estimator before moving forward.

\subsection{Geometric Partition Mutual Information}
We will rely on the Partition Intersection ($\pi$) estimator of GPMI with the chosen measure being the product of observation frequency ($f$) and a geometric quantity ($g$), e.g., the Euclidean norm.  This approach to information-theoretic analysis was specifically designed with transient comparisons during synchronization in mind, and the quantile-based estimator ($I_\pi$) for GPMI can be computed several orders of magnitude faster.  It is also more data-efficient and enables the use of shorter trajectories.  The ability of GPMI to leverage different choices of measure also enables dynamic weighting of the transient period when warranted, e.g., a specific volume measure (inverse density) enables a significant differentiation between transient dynamics similar to synchronization error, but in terms of information theoretic measures.  

Our chosen measure throughout this paper, i.e. $m(f,g)=f\cdot g$, provides a consistent and balanced weighting of the synchronous manifold and transient portions of the dynamics, which excels in the current application, where we are interested in both synchronization generally, but also the time to synchronization. Again, the actual value of the GPMI estimate will be highly dependent on the choice in measure, but our choice here is remarkably consistent in variation with KSG estimates (and consistently scales closely with histogram estimates).

\subsection{Network Flow Maps}
To compliment our use of MI heatmaps, we also provide a network-based measure of Conditional Mutual Information (CMI) based on the principle of CauSation Entropy (CSE)~\cite{sun2014causation}.  Using the known underlying network structure, we consider only those pairs connected by a link in the network; for each such pair, we compute (in both directions) the mutual information between the target node dynamics and a one-step time delay of the source node dynamics, conditioned on the time-delayed dynamics for the target node and all other adjacent nodes in the network. Here, we use the same $I_\pi$ estimator as before.  In the spirit of Transfer Entropy and CSE, this particular choice of measure can be viewed as a quantification of ``information flow''; however, the analogy to flow is not strictly correct as was demonstrated in~\cite{james2016information}.

These CMI measures quantify the information that the source history provides about the target's future given the conditioning set. They can combine information unique to the source with synergistic information available only through its joint contribution with the conditioning variables, without separately resolving these contributions. This limits their interpretation as information flowing from the source to the target. We nevertheless refer to the resulting plots as ``network flow heatmaps'', with this qualification understood.

Importantly, these network flow heatmaps are not symmetric, and the larger value when compared with the opposite direction can indicate an increased influence in one direction, i.e. net flow.  However, when using these plots, it is also important to ``take into consideration'' the size of the conditioning set in each case, since this changes based on the degree of the target node and makes a significant difference in the overall value. This means that unlike in the case of MI, comparisons of CMI values are not necessarily direct and must consider the degree of the target node involved.

Additional confusion with these plots may arise from the fact that when neighboring nodes synchronize quickly, the ``information flow'' between these near synchronous nodes goes toward zero rapidly as the contribution from the coupling term vanishes.  On the other extreme, large values of CMI can stem from either being a potential bottleneck link that has a long transient due to facilitating the synchronous state or an indication of a total failure to synchronize, and we will see both cases.  Thus, in the low coupling strength regime, where $\sigma$ is chosen near the MSF-defined threshold, high value CMI measures will indicate a higher attempt to influence neighboring nodes over the bottleneck edges; but, they do not necessarily indicate a failure to synchronize, and the changes in node degree must always be kept in mind. 

\subsection{Results}
All information measure estimates are averaged over $r=100$ sets of initial conditions, which are randomized Gaussian perturbations from the synchronous state (within the basin of stability), each continuous trajectory is numerically integrated using \textit{odeint} over a time interval $t\in[0,T]$ for an integer $T$ with a fixed step size of $\Delta t=1/10$ in all cases. We use a parameter value of $k=20$ nearest-neighbors for the KSG estimator and $k=10$ quantiles-per-dimension for the $I_\pi$ estimator of GPMI. Since we use the same measure for $I_\pi$ across all simulations, the only parameters that differ between plots are the global coupling strength $\sigma$ and the total time $T$, which ranges from $T=5000$ for smaller graphs to $T=10,000$ for larger graphs that exhibit longer transient dynamics.

\subsubsection{Small-set Bottleneck Identification}
The graph in Fig.~\ref{fig:defns} provides an unweighted example where more than one bottleneck is present.  Although the conductance of the graph $G$ was measured as $\Phi_G=1/18$, the minimum cut, $C$, that is associated with the conductance has a synchronization bottleneck ratio of $\Xi(G,C)=1/3$, which is larger than the value of $\Xi_G=2/9$, though not significantly so. Choosing $\sigma$ to ensure that the coupling strength-boosted eigenvalue $\sigma\lambda_2<K_\alpha=0.08$ (e.g., $\sigma=3/5$ resulting in $K\approx 0.068$) leads to a failure to synchronize; while on the other hand, a choice of $\sigma=3/4$ places us in the complete synchrony regime with $K\approx0.085>K_\alpha$.  

\begin{figure}[ht!]
\begin{tabular}{>{\centering\arraybackslash}m{0.4cm}>{\centering\arraybackslash}m{3.75cm}>{\centering\arraybackslash}m{3.75cm}}
\rotatebox{90}{\qquad\qquad KSG}&\includegraphics[width=3.75cm]{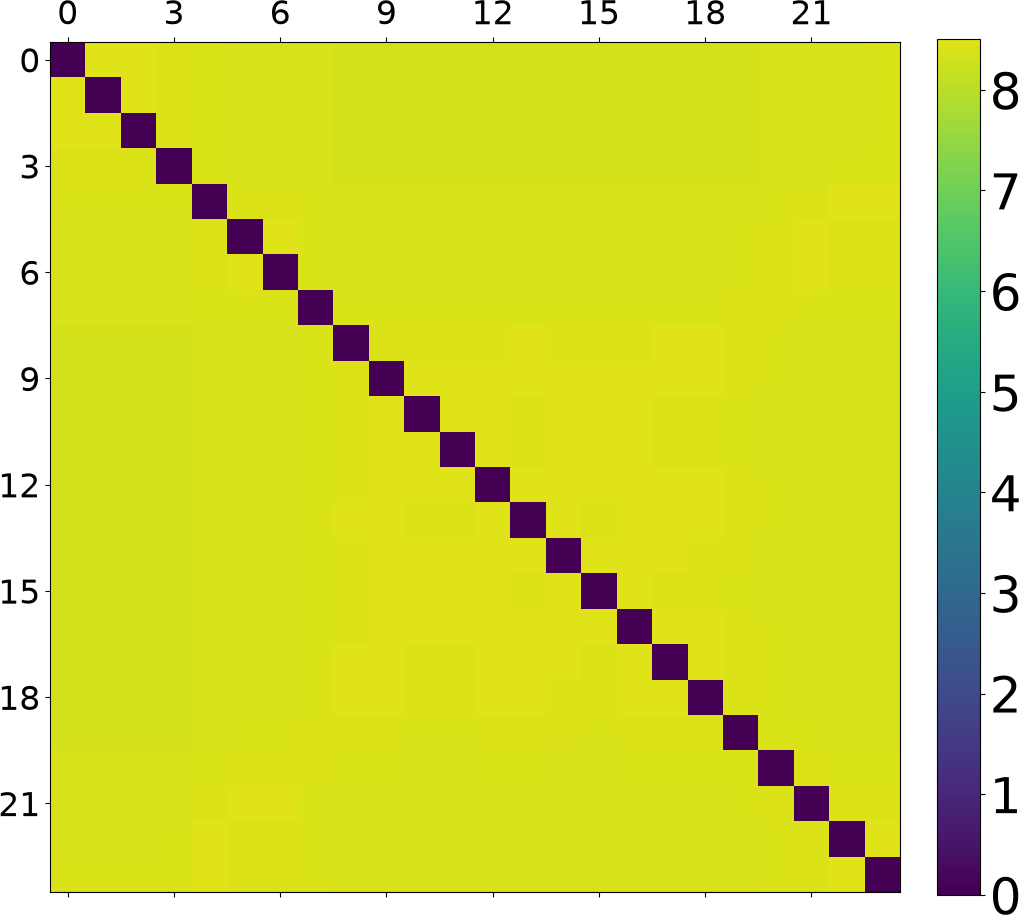}&\includegraphics[width=3.75cm]{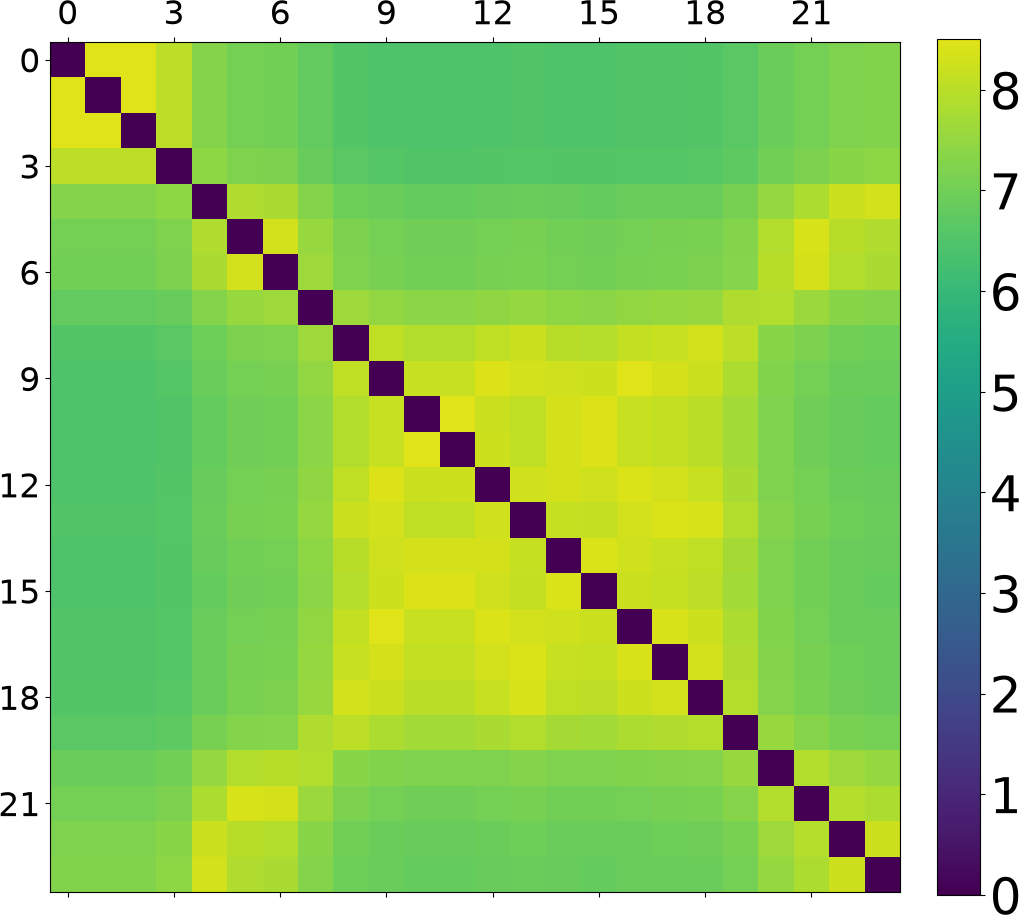}\\
&$\sigma=3/4$ & $\sigma=3/5$\\
&(a)& (b)\\[0.1cm]
\hline\\[0.1cm]
\rotatebox{90}{\qquad\qquad \quad $I_\pi$}&\includegraphics[width=3.75cm]{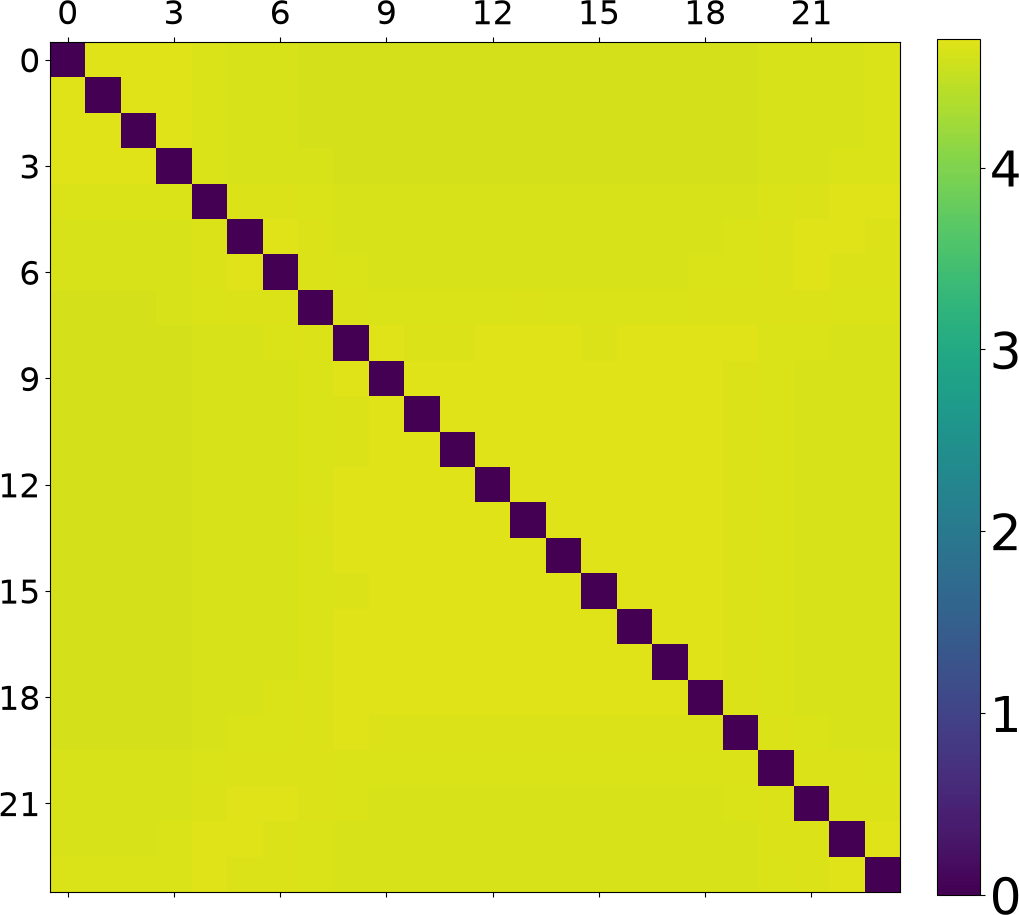}&
\includegraphics[width=3.75cm]{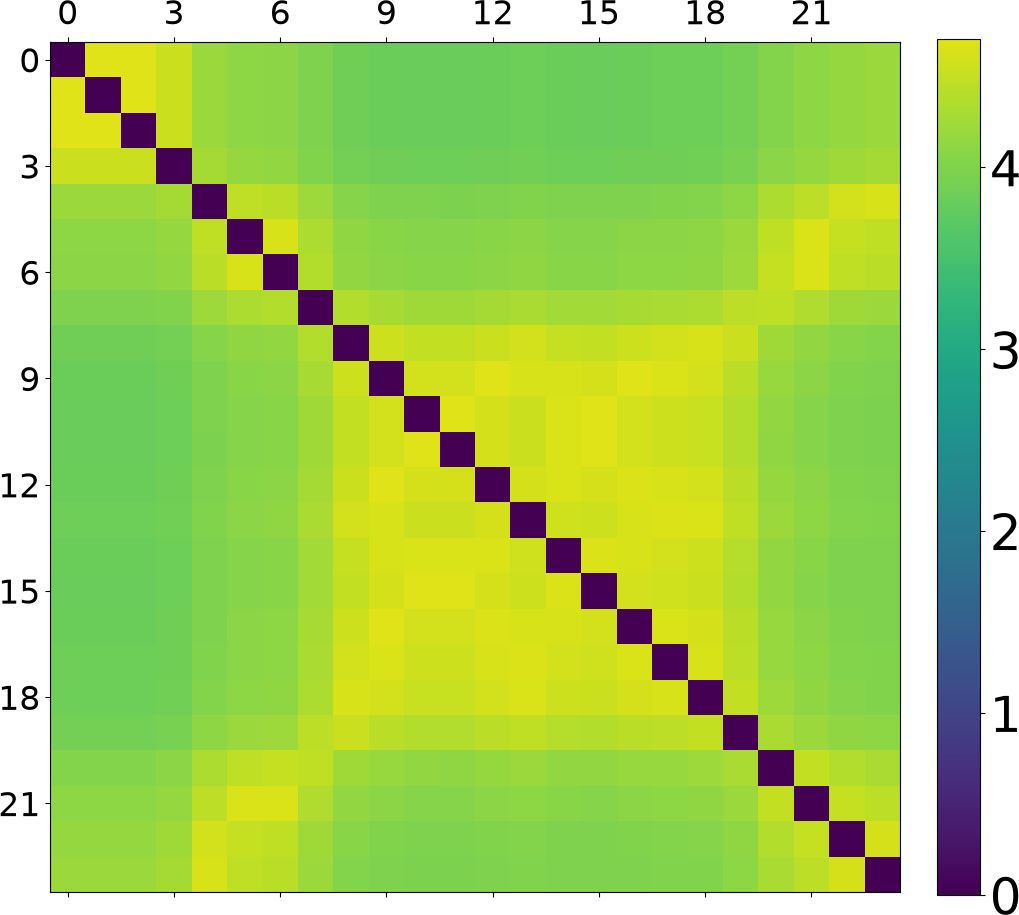}\\
&$\sigma=3/4$ & $\sigma=3/5$\\
&(c)& (d)\\
\end{tabular}
\caption{MI heatmaps for $\sigma\lambda>K_\alpha$ in (a) and (c) and $\sigma\lambda<K_\alpha$ in (b) and (d); measures were estimated by the KSG estimator in (a) and (b) and the $I_\pi$ estimator with $m(f,g)=f\cdot g$ in (c) and (d); although the scale is different between the two estimators, the qualitative comparisons are almost identical, supporting our use of GPMI.\label{fig:bottleneckMI}}
\end{figure}

First, applying the KSG estimator with $\sigma=3/4$ to all pairs of trajectories (with $T=10,000$) we obtain the plot in Fig.~\ref{fig:bottleneckMI} (a) indicating complete synchrony. Changing the global coupling strength to $\sigma=3/5$ results in the expected failure to synchronize, which is clear from the heterogeneity of the MI heatmap in (b).  The KSG estimates serve as a reliable and trusted control enabling a validation of GPMI~\cite{diggans2025generalizing} by providing these same MI heatmap plots as estimated by the $I_\pi$ estimator in (c) and (d).  Although the scale of measure is different, the $I_\pi$ estimates are actually closer to traditional histogram estimates and as we stated above, the actual values are dependent on parameters and thus unimportant.  The relevant feature is the comparison across similarly applied estimates, and here we see complete consistence with the KSG estimator.  Since these estimates are obtained at a fraction of the computational time, the GPMI estimator was used throughout all remaining plots.

\begin{figure}[ht!]
\begin{tabular}{>{\centering\arraybackslash}m{4cm}>{\centering\arraybackslash}m{4cm}}
\includegraphics[width=4cm]{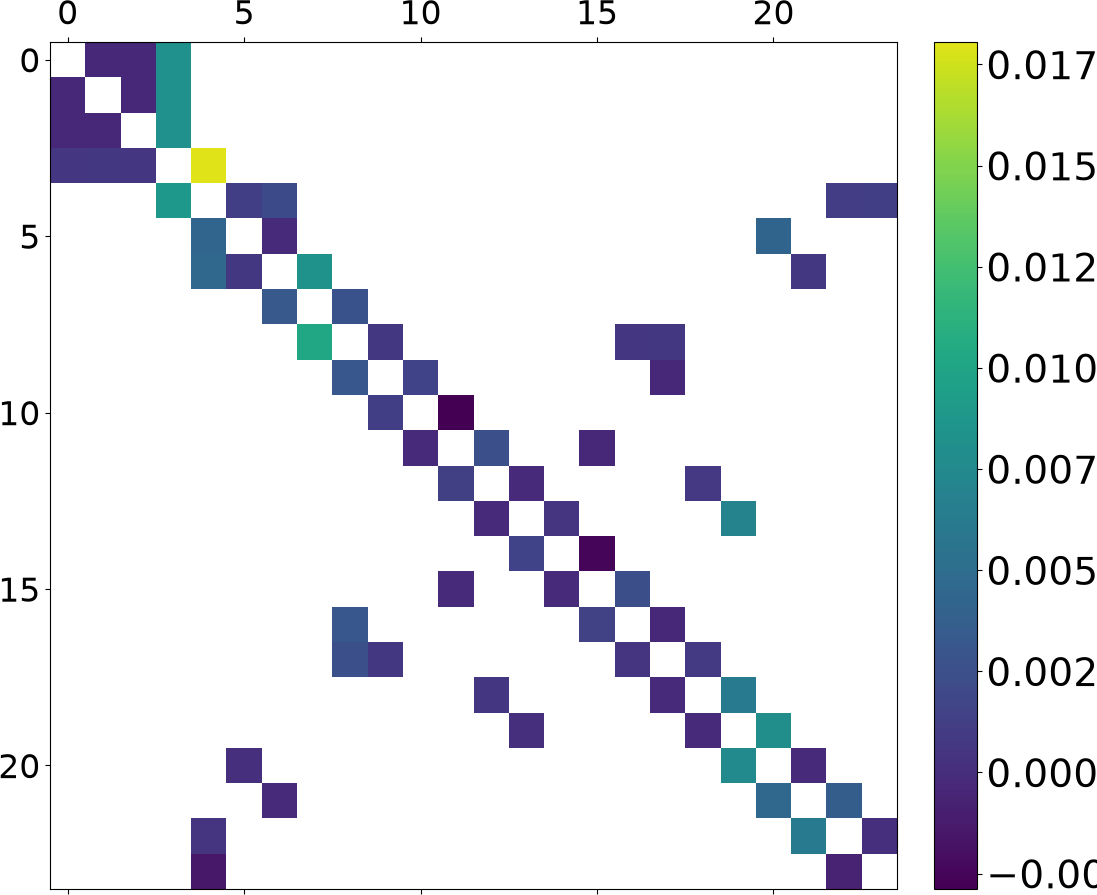}&
\includegraphics[width=4cm]{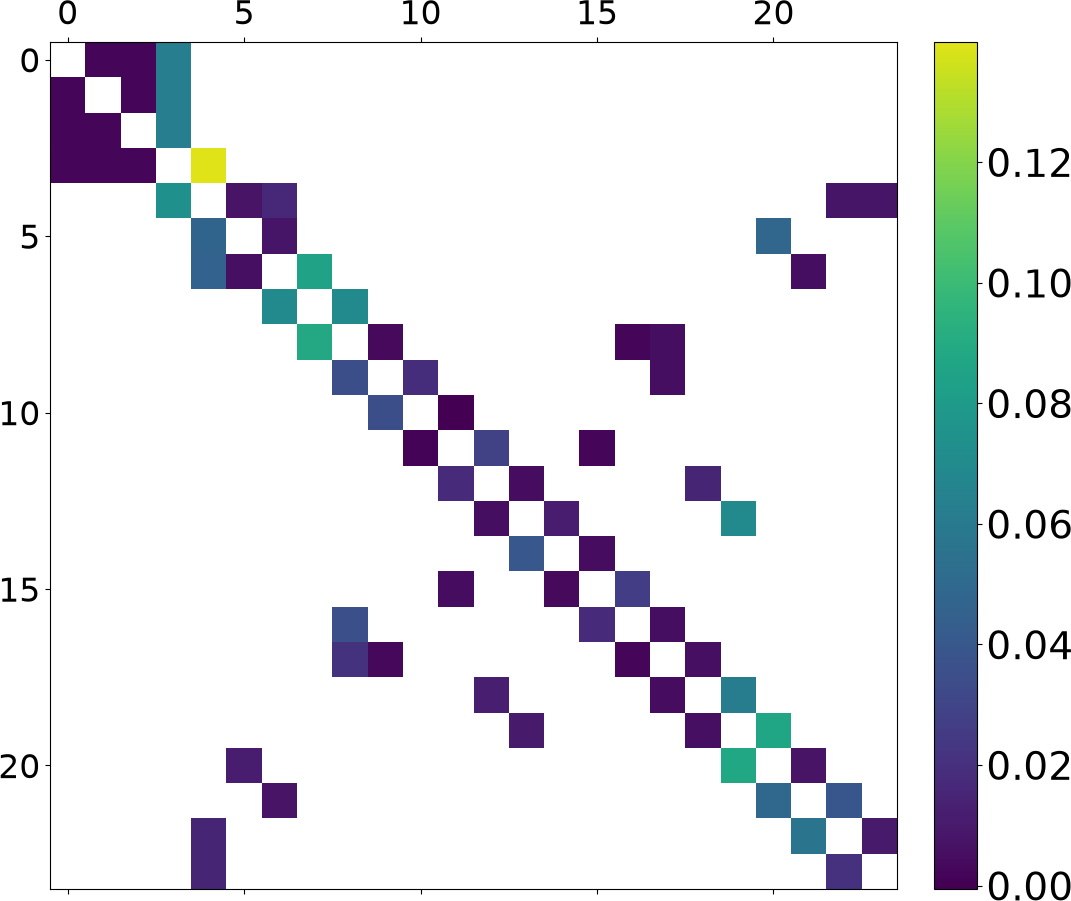}\\
$\sigma=3/4$ & $\sigma=3/5$\\
(a)& (b)\\
\end{tabular}
\caption{The CSE-inspired network flow heatmaps estimated using $I_\pi$ estimator with $m(f,g)=f\cdot g$ for (a) $\sigma=3/4$ and (b) $\sigma=3/5$; although the general relationships are similar, note a difference in the role played by node $7$ and, most importantly, the difference in scale.\label{fig:bottleneckCMI}}
\end{figure}

In both (b) and (d), we see the same evidence that both relevant cuts are impacting synchronization.  Comparisons of MI by themselves may not differentiate between longer transient dynamics and a true failure to synchronize without variation in the parameter $T$, but we see additional evidence that the small-set bottleneck is driving the failure to synchronize in this case as we consider the CSE-inspired network flow maps in Figure~\ref{fig:bottleneckCMI} (a) and (b).  We only show the $I_\pi$ estimates, but they are again consistent with KSG estimates, even with respect to the appearance of small negative values; and here the scale even matches in both cases since we are measuring CMI.

The first thing to notice when comparing Fig.~\ref{fig:bottleneckCMI} (a) and (b) is the change in scale, which indicates the clearest difference between stable synchronous state and a configuration that can not achieve complete synchrony.  Recalling that the degrees of the nodes should be taken into consideration when comparing the CSE values across an individual plot, an excellent example is found here by considering node $7$. In (a) we can see that there is a clear difference in color indicating variation in the flow going in-to and out-from node $7$ with each of its neighbors.  When the dynamics are synchronous, this difference is explained by the differing degrees, since $deg(7)=2$ while $deg(6)=4=deg(8)$.  However, comparing this difference in value with what we see in (b), where the graph no longer synchronizes, we see that even though the CSE values are more symmetric, this indicate an observation of imbalance in flow, which is either an indication of a role in the leadership for a synchronous state or a failure to synchronize altogether.  The same indication that the overall scale provides for a failure to synchronize also indicates here that the small-set bottleneck may in fact be the true limiting bottleneck; despite the high degrees of both nodes adjacent to the cut link, we find a significantly larger CSE measure. 

Another indication of this small-set cut being the limiting bottleneck is that we find the constant pull from node $4$ on the state of node $3$ results in an increased influence from node $3$ on the rest of the small-set, nodes $\left\{0,1,2\right\}$, as indicated by the lighter color in the column associated with node $3$; this phenomenon is not apparent when the neighboring nodes of $4$ are considered. The difference in the directional flow measurements on $\left\{3,4\right\}$ is in part driven by a difference in degree again, but it is also due to the difference in the sizes of the connected components on either side of the cut. Finally, we observe that the other link involved in the minimum cut for conductance, i.e. $\left\{19,20\right\}$, has not changed in a noticeable way in relation to its neighbors.  This then means that the increased flow of information coming to node $7$ from both of its neighbors indicates that its dynamics are getting frustrated by two opposing pulls; an indication that perhaps the larger connected component would still be able to synchronize if the small-set bottleneck was not consistently pulling on the state of node $4$ in greater quantaties.  This further indicates that the small-set bottleneck is in fact the likely limiting bottleneck in this case. Each of these cases of complex networks is a challenge to make sense of, so we now turn to the simplified reduced order model.

\subsubsection{Verifying the Predicted Tipping Points}
\begin{figure}[ht!]
\centering
\begin{tabular}{>{\centering\arraybackslash}m{0.4cm}>{\centering\arraybackslash}m{0.4cm}|>{\centering\arraybackslash}m{2.5cm}|>{\centering\arraybackslash}m{2.75cm}}
&&$MI_\pi$& $CSE_\pi$\\
\hline
&&\multicolumn{2}{c|}{Synchronized (Control)}\\
\hline
\rotatebox{90}{\qquad$\alpha=1/2$}&\rotatebox{90}{\qquad$\sigma=0.30$}&\rule{0pt}{2pt} \includegraphics[width=2.5cm]{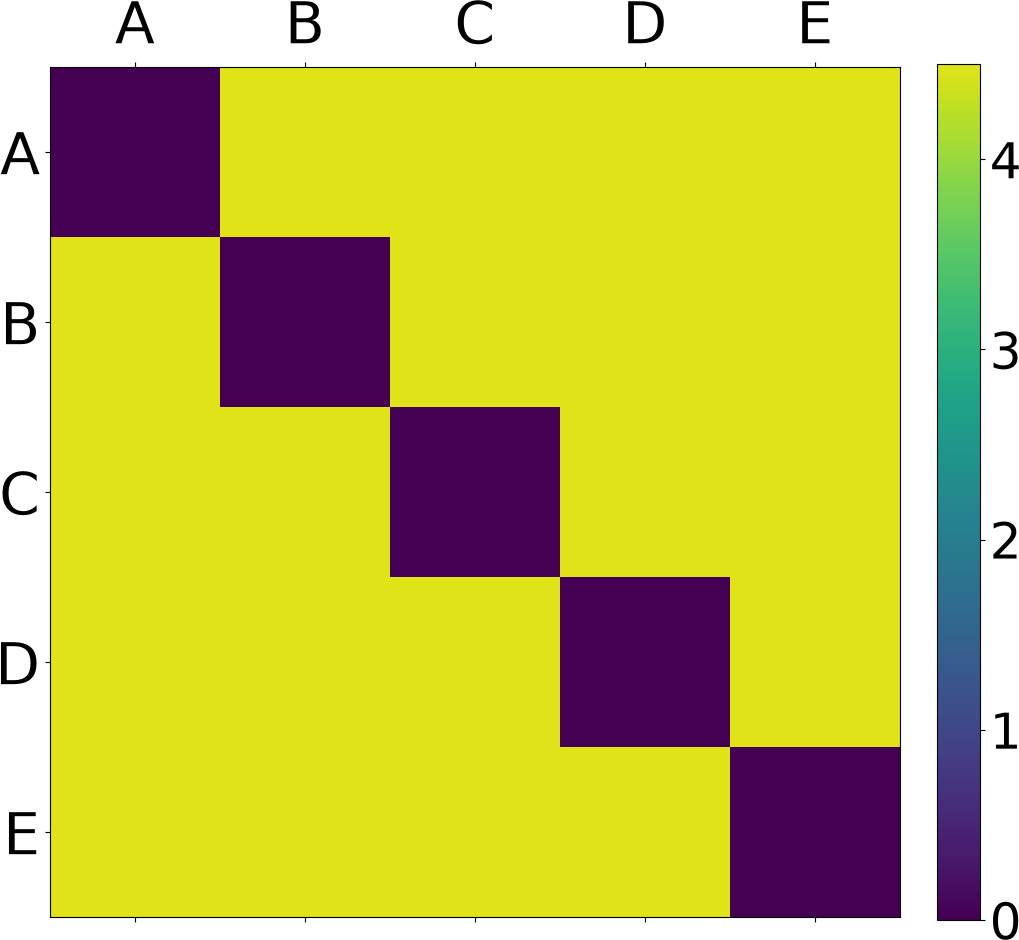} &\rule{0pt}{2pt} \includegraphics[width=2.65cm]{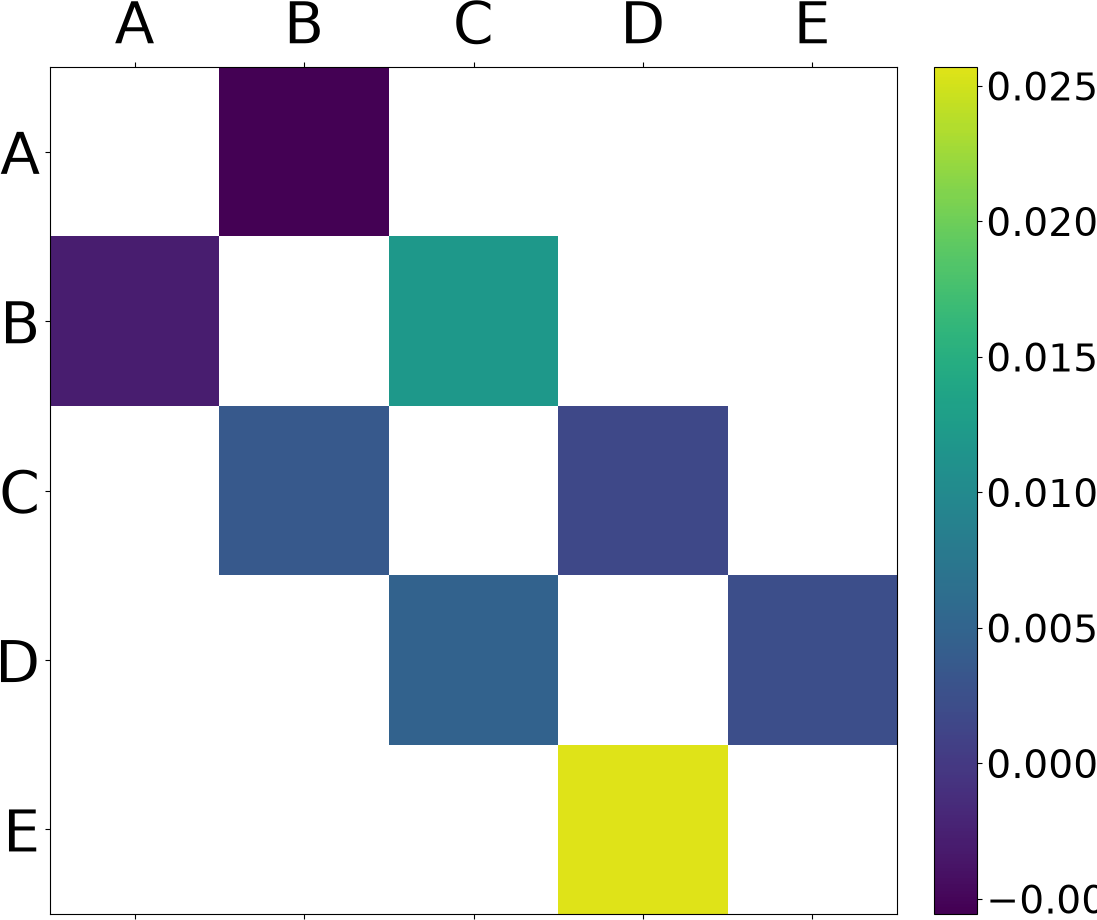}\\
\hline
&&\multicolumn{2}{c|}{Chimera States}\\
\hline
\rotatebox{90}{\qquad$\alpha=1/4$}&\rotatebox{90}{\qquad$\sigma=0.250$}&\rule{0pt}{2pt} \includegraphics[width=2.5cm]{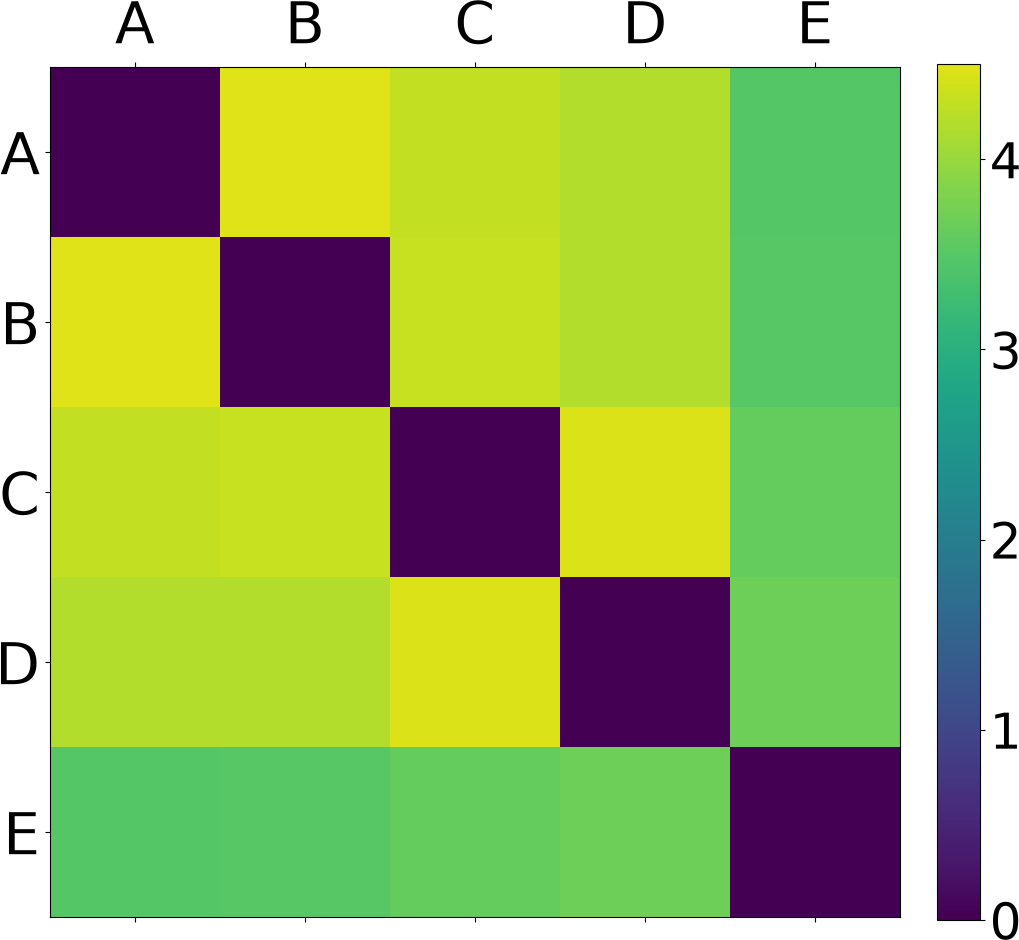}&\rule{0pt}{2pt} \includegraphics[width=2.65cm]{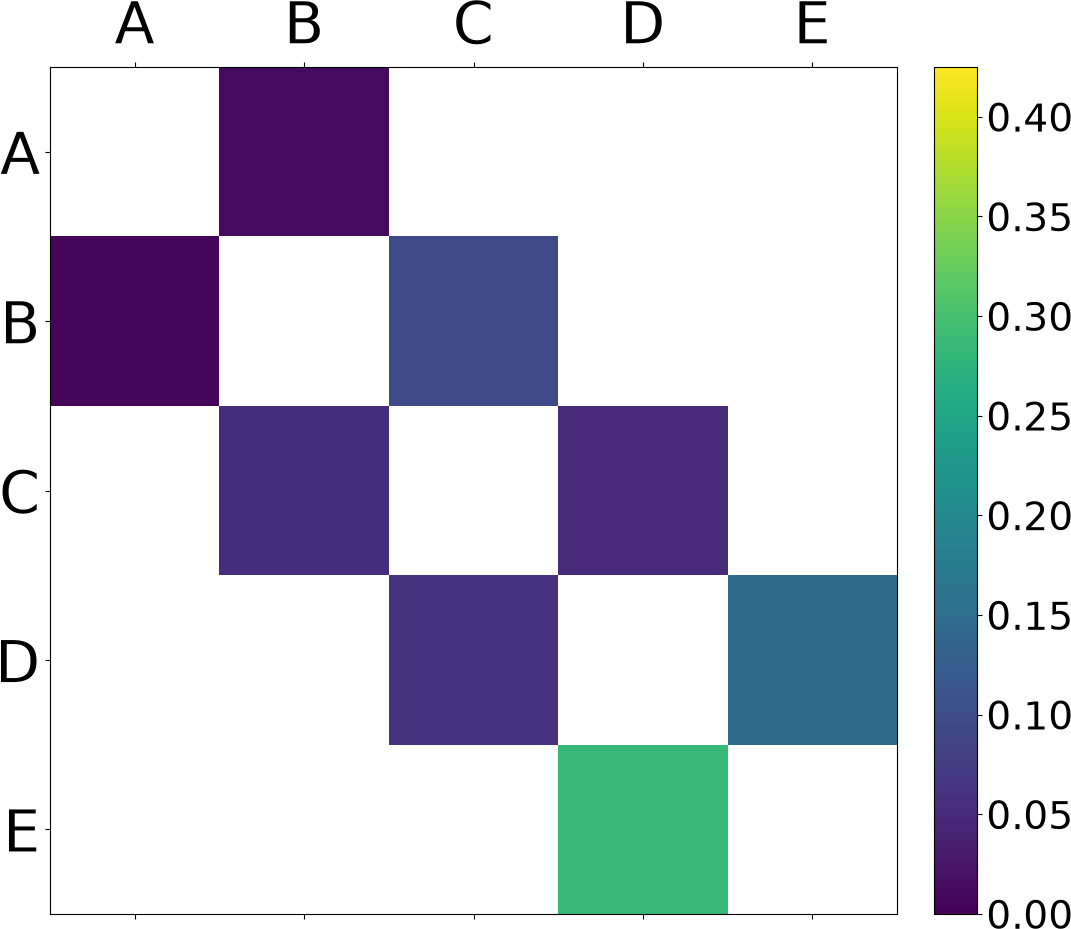}\\
\multicolumn{2}{c|}{(a)}& (i)& (ii)\\[4pt]
\hline
\rotatebox{90}{\qquad$\alpha=1/3$}&\rotatebox{90}{\qquad$\sigma=0.200$}&\rule{0pt}{2pt} \includegraphics[width=2.5cm]{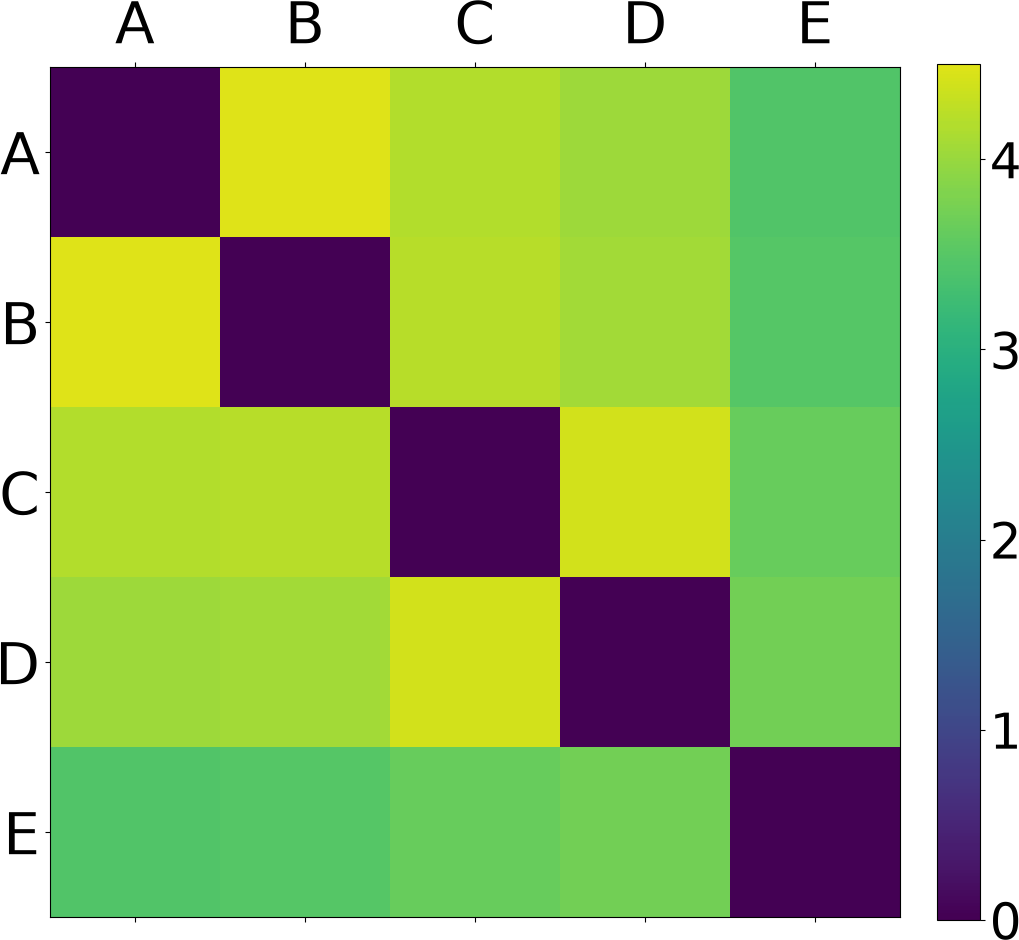}&\rule{0pt}{2pt} \includegraphics[width=2.65cm]{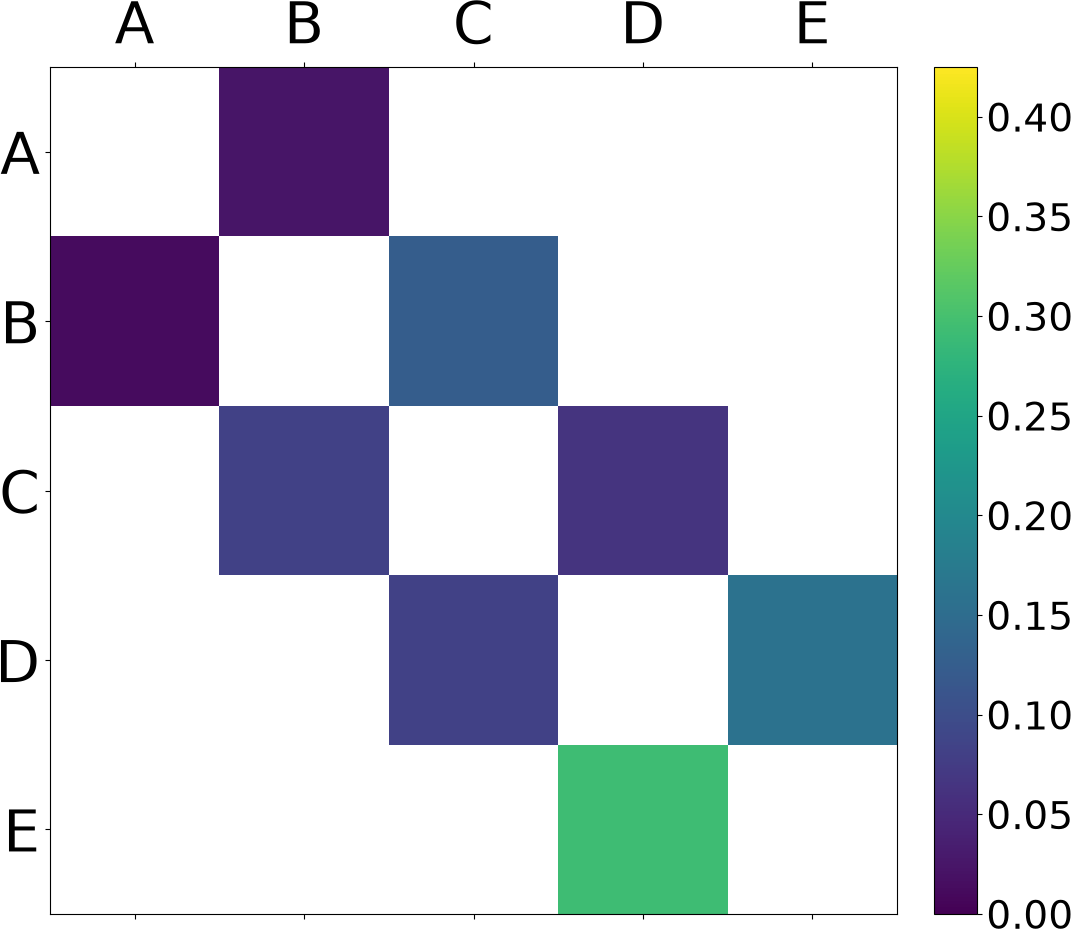}\\
\multicolumn{2}{c|}{(b)} & (i)& (ii) \\[4pt]
\hline
\rotatebox{90}{\qquad$\alpha=1/2$}&\rotatebox{90}{\qquad$\sigma=0.155$}&\rule{0pt}{2pt} \includegraphics[width=2.5cm]{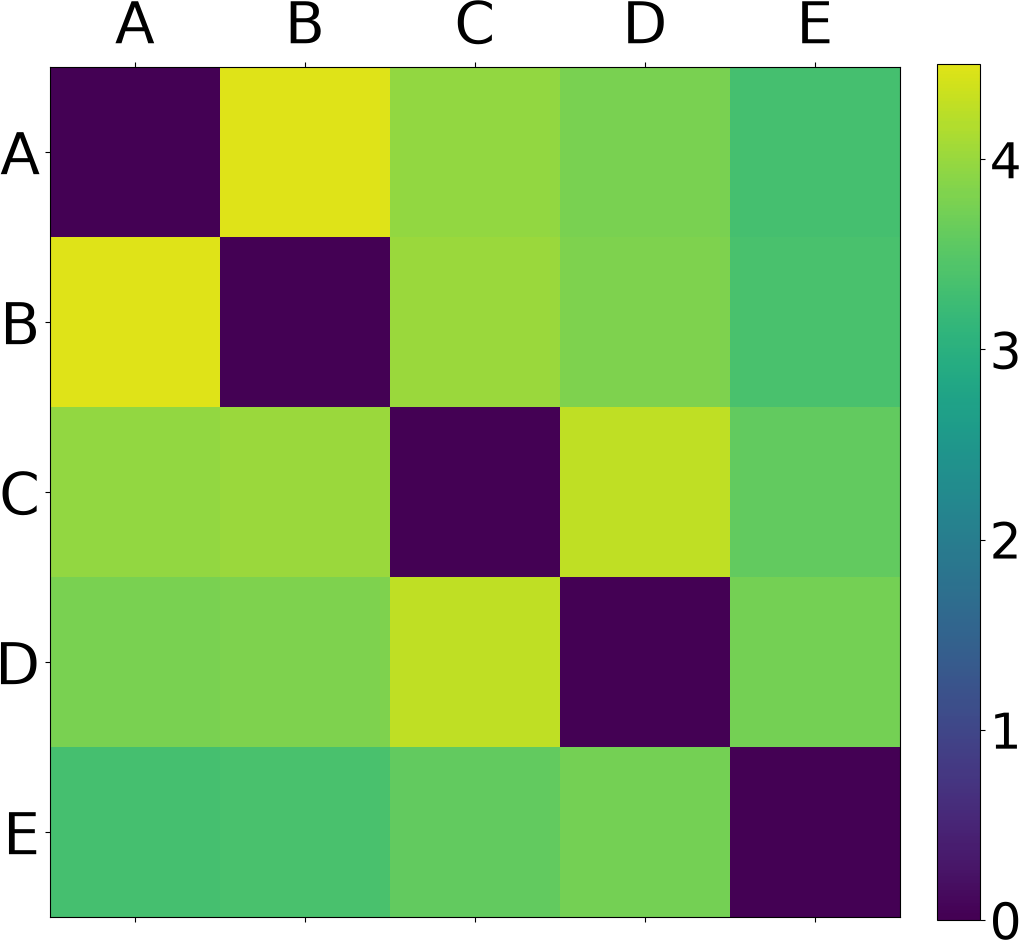}&\rule{0pt}{2pt} \includegraphics[width=2.65cm]{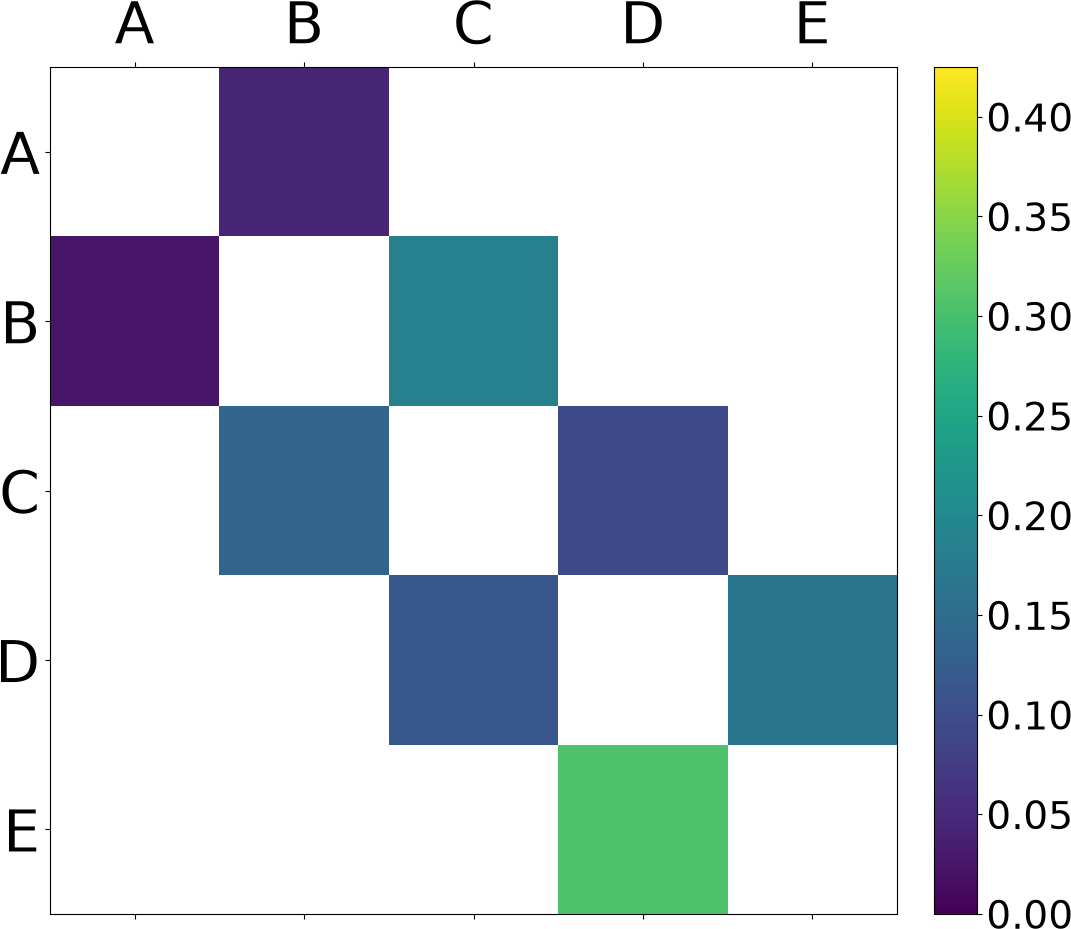}\\
\multicolumn{2}{c|}{(c)} & (i)& (ii) \\[4pt]
\hline
\rotatebox{90}{\qquad$\alpha=3/4$}&\rotatebox{90}{\qquad$\sigma=0.135$}&\rule{0pt}{4pt} \includegraphics[width=2.5cm]{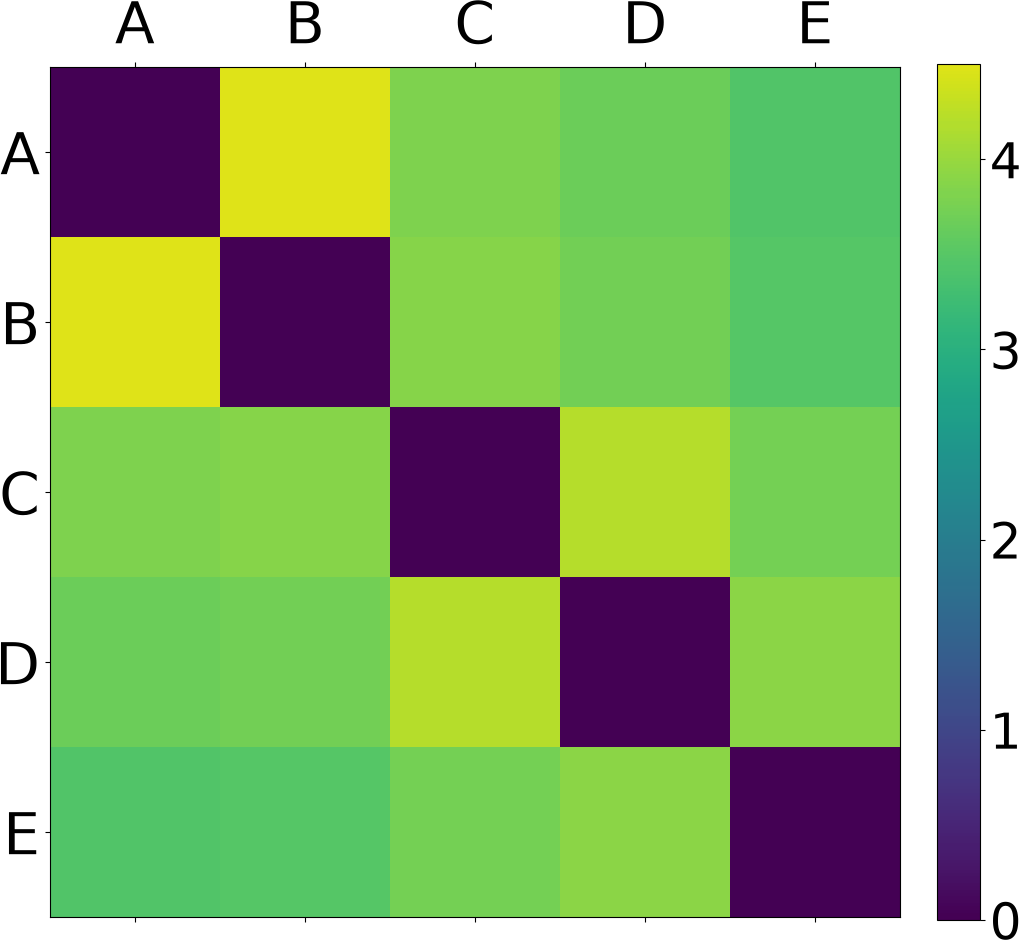}&\rule{0pt}{4pt} \includegraphics[width=2.65cm]{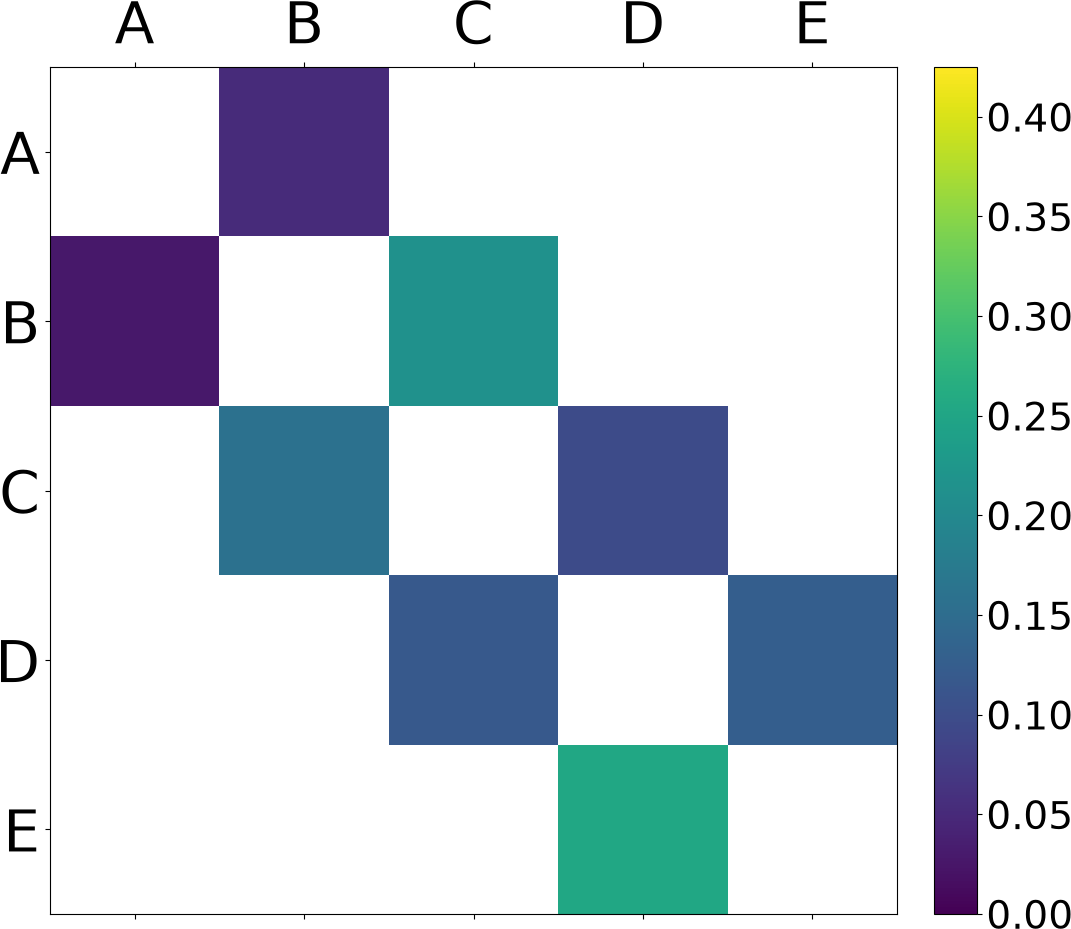}\\
\multicolumn{2}{c|}{(d)} & (i)& (ii) \\[4pt]
\hline
\rotatebox{90}{\qquad$\alpha=1$}&\rotatebox{90}{\quad$\sigma=0.125$}&\rule{0pt}{5pt} \includegraphics[width=2.5cm]{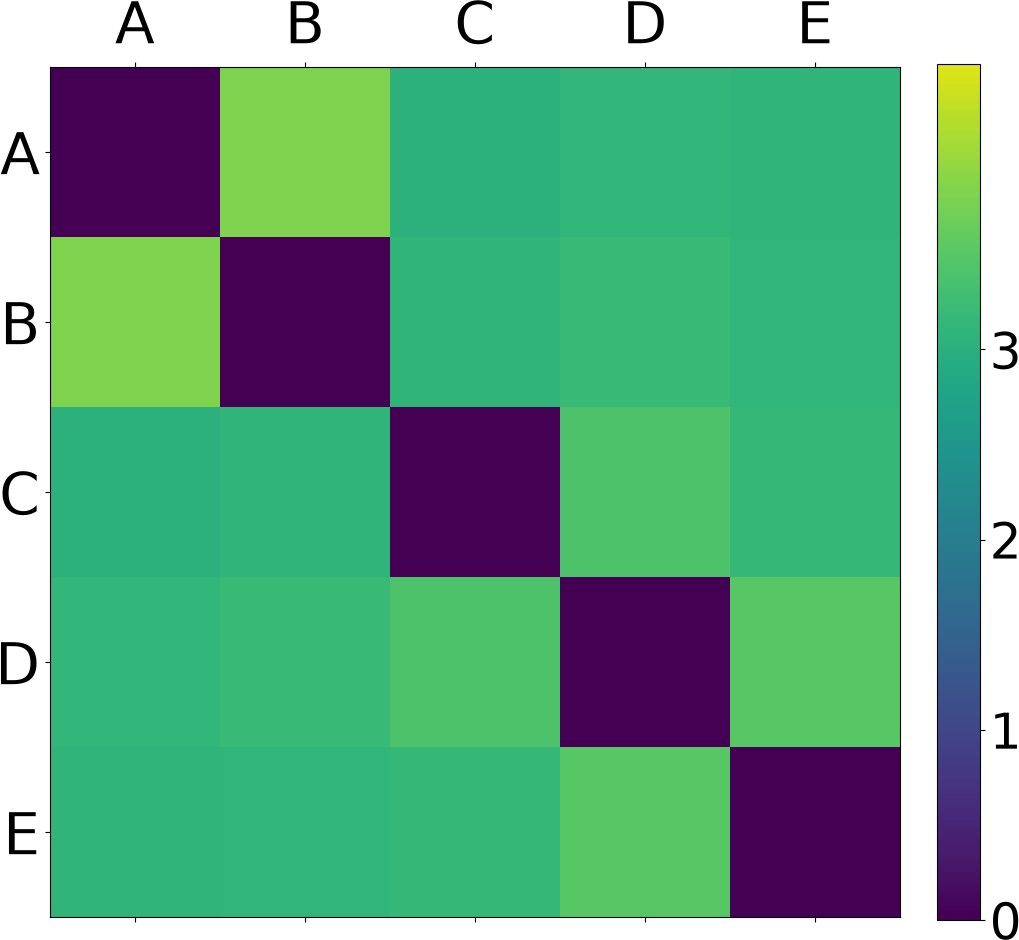}&\rule{0pt}{4pt} \includegraphics[width=2.65cm]{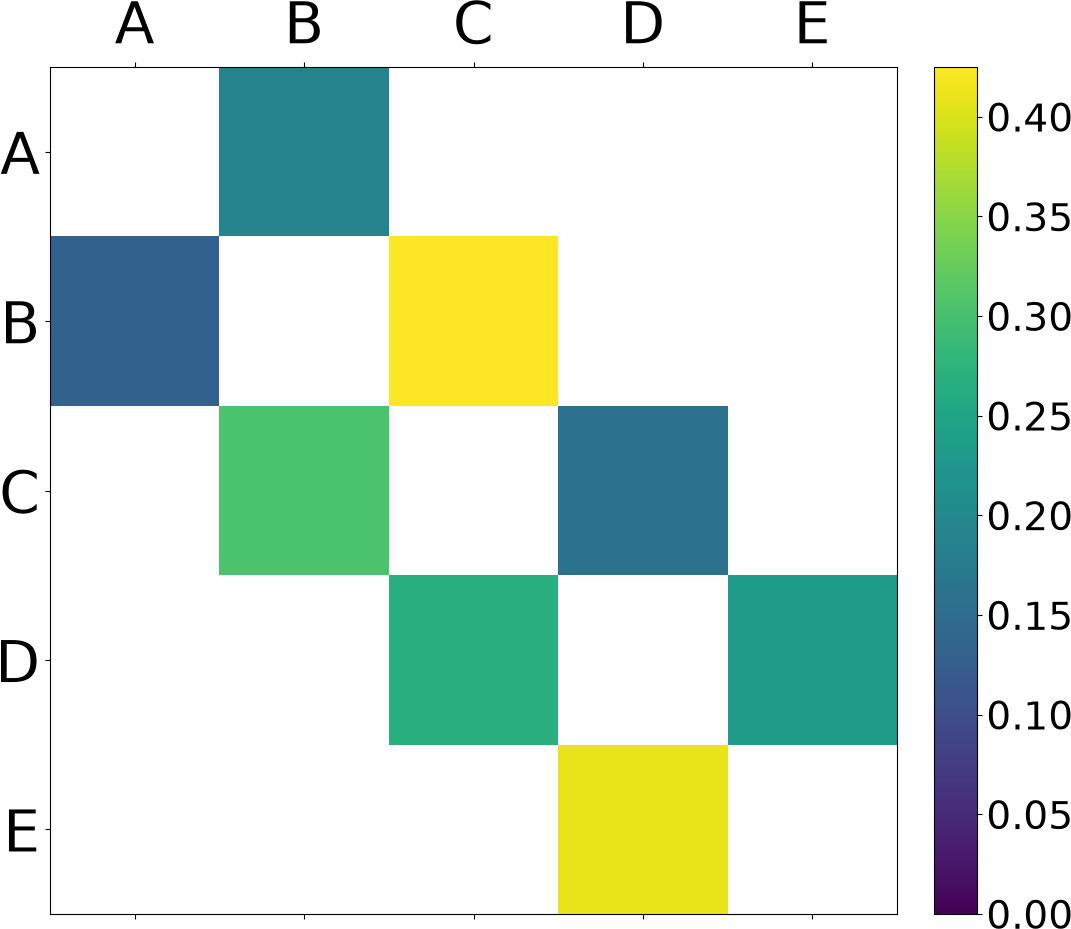}\\
\multicolumn{2}{c|}{(e)} & (i)& (ii) \\[4pt]
\end{tabular}
\caption{Based on trajectories of identical R\"{o}ssler attractors coupled on the weighted path graph with $w_\ell=4$ and $w_c=2$ held fixed for the values of $\alpha$ and $\sigma$ indicated in (a)-(e); where for each set of parameters, we provide the i) GPMI and ii) network flow heatmaps using the $I_pi$ estimator with $m(f,g)=f\cdot g$ for $T=10,000$.\label{fig:P5_MI}}
\end{figure}

Considering the weighted path graph from before, we now explore the transition point, i.e. tipping point, as the parameterized $\alpha$-pendant increases in edge weight. Figure~\ref{fig:P5_MI} shows plots of the GPMI for the weighted path graph from Fig.~\ref{P5}, where we hold $w_\ell=4$ and $w_c=2$ fixed as before. In comparison to the control case case, where the dynamics reaches a synchronized state, we consider ranging $\alpha$ over five different values in (a)-(e), where in each case, $\sigma$ has been chosen to produce the same value for the $\sigma$-boosted eigenvalue just below the MSF-threshold (around $K\approx 0.065$) to ensure similar information-based indicators.  The plots in the first column (i), indicate a gradual transition in the limiting cut, from the pendant cut $\left\{D,E\right\}$ in (a) to the cut $\left\{B,C\right\}$ overtaking that pendant cut clearly in (e), where the nodes $\left\{C,D,E\right\}$ are beginning to share more information.  That transition is made clearer through the CSE network flow maps, where the change in the maximum flow estimate is clear by the time $\alpha=1$.  Emphasizing, yet again, that the weighted degrees of the nodes play an important role in this measure, it is helpful to compare each case to the synchronized control provided for $\alpha=1/2$. Although the scale of values is considerably reduced, the relationships between the states provide a control case with respect to the impact that varies with degree.  And although identifying the exact tipping point is confounded by several additional factors that impact the information measures, including variations in transient dynamics and the chosen simulation time, the direct comparison with the control case indicates that the pendant vertex is still the limiting cut in (c). 

So, despite the benefit as an approximation, it is not likely that our measure provides a robust estimate of the true limiting bottleneck.

Finally we turn to the Ramanujan---$\alpha$ pendant graph, which is an extreme case where the Fiedler subspace is more complex having many redundancies.  We still observe the correct limiting bottleneck being identified for small values of $\alpha$, and although it is not clear whether longer simulation times would prove eventual convergence for larger values of $\alpha$, we do also observe a clear change in structure between $\alpha=1$ and $\alpha=5/4$, exactly where our measure predicts a tipping point. But, as $\alpha$ grows larger, e.g. $\alpha=5$, the global coupling strength must be reduced significantly to produce chimera states, and it appears that synchronization breaks down all over the network rather than at any particular cut due to the high redundancy in the Ramanujan graph.  
\begin{figure}[ht!]
\centering
\begin{tabular}{>{\centering\arraybackslash}m{0.4cm}>{\centering\arraybackslash}m{0.4cm}|>{\centering\arraybackslash}m{3.5cm}|>{\centering\arraybackslash}m{3.5cm}}
&&$MI_\pi$& $CSE_\pi$\\
\hline
\rotatebox{90}{\qquad$\alpha=1$}&\rotatebox{90}{\qquad$\sigma=0.075$}&\rule{0pt}{3pt} \includegraphics[width=3.5cm]{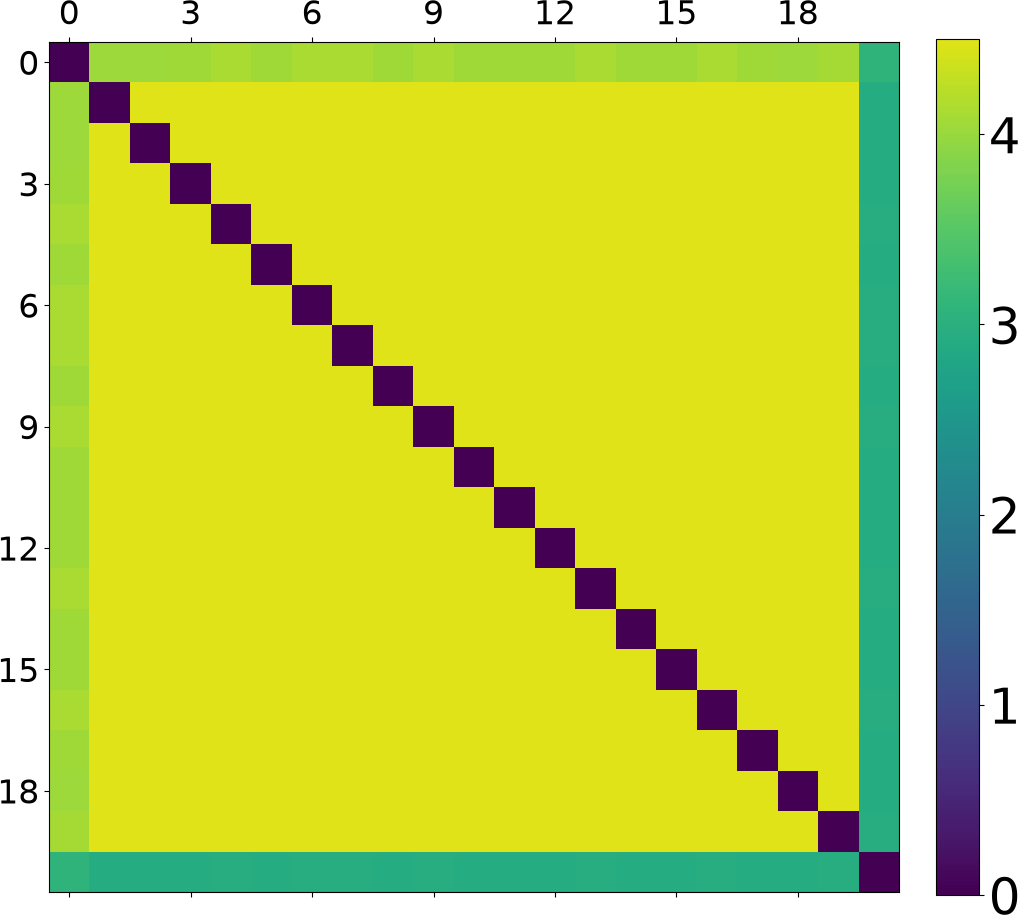}&\rule{0pt}{3pt}\includegraphics[width=3.5cm]{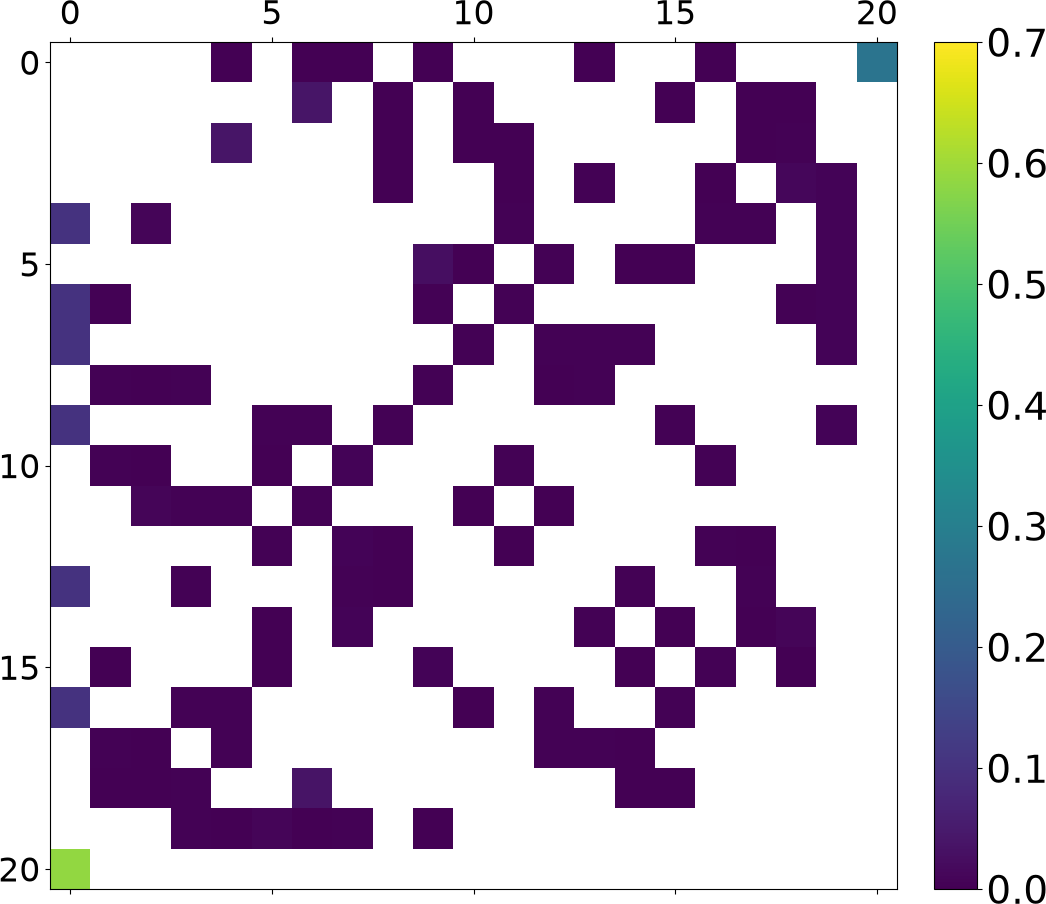}\\
\multicolumn{2}{c|}{(a)}& (i)& (ii)\\[2pt]
\hline
\rotatebox{90}{\qquad$\alpha=5/4$}&\rotatebox{90}{\qquad$\sigma=0.062$}&\rule{0pt}{3pt} \includegraphics[width=3.5cm]{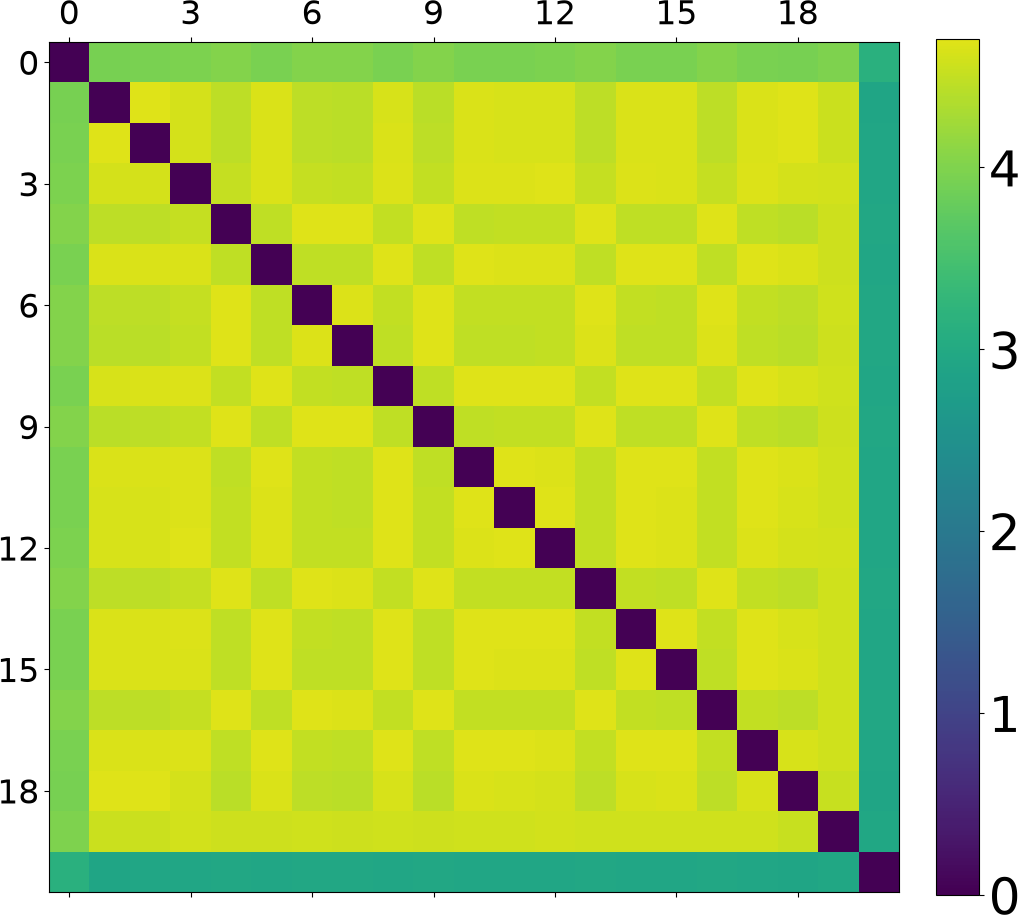}&\rule{0pt}{3pt} \includegraphics[width=3.5cm]{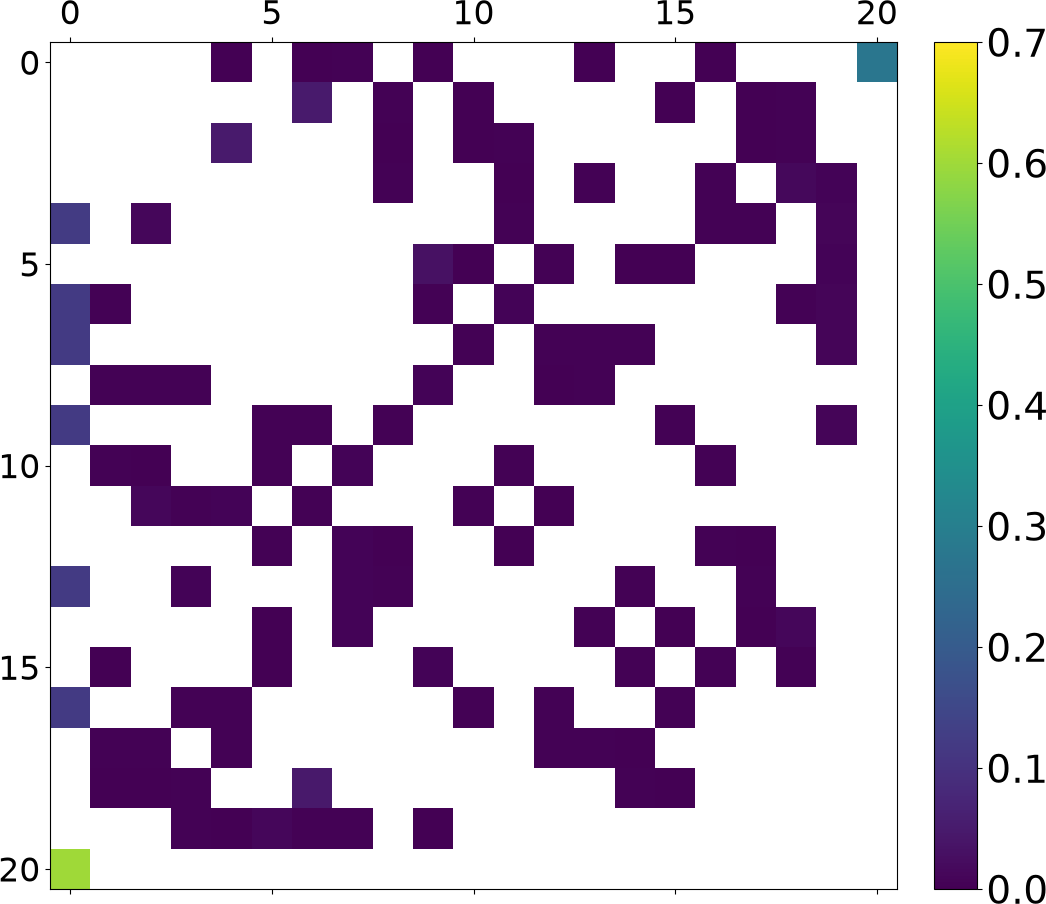}\\
\multicolumn{2}{c|}{(b)}& (i)& (ii)\\[2pt]
\hline
\rotatebox{90}{\qquad$\alpha=2$}&\rotatebox{90}{\qquad$\sigma=0.045$}&\rule{0pt}{3pt} \includegraphics[width=3.5cm]{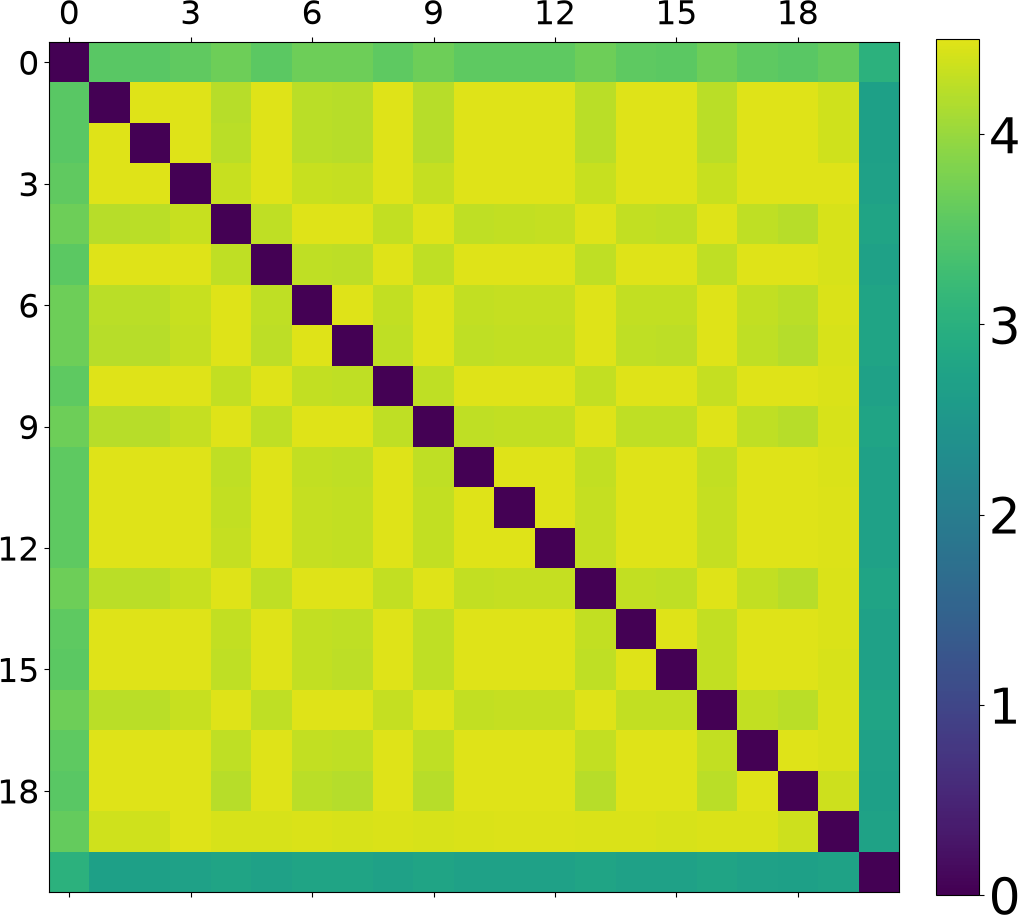}&\rule{0pt}{3pt} \includegraphics[width=3.5cm]{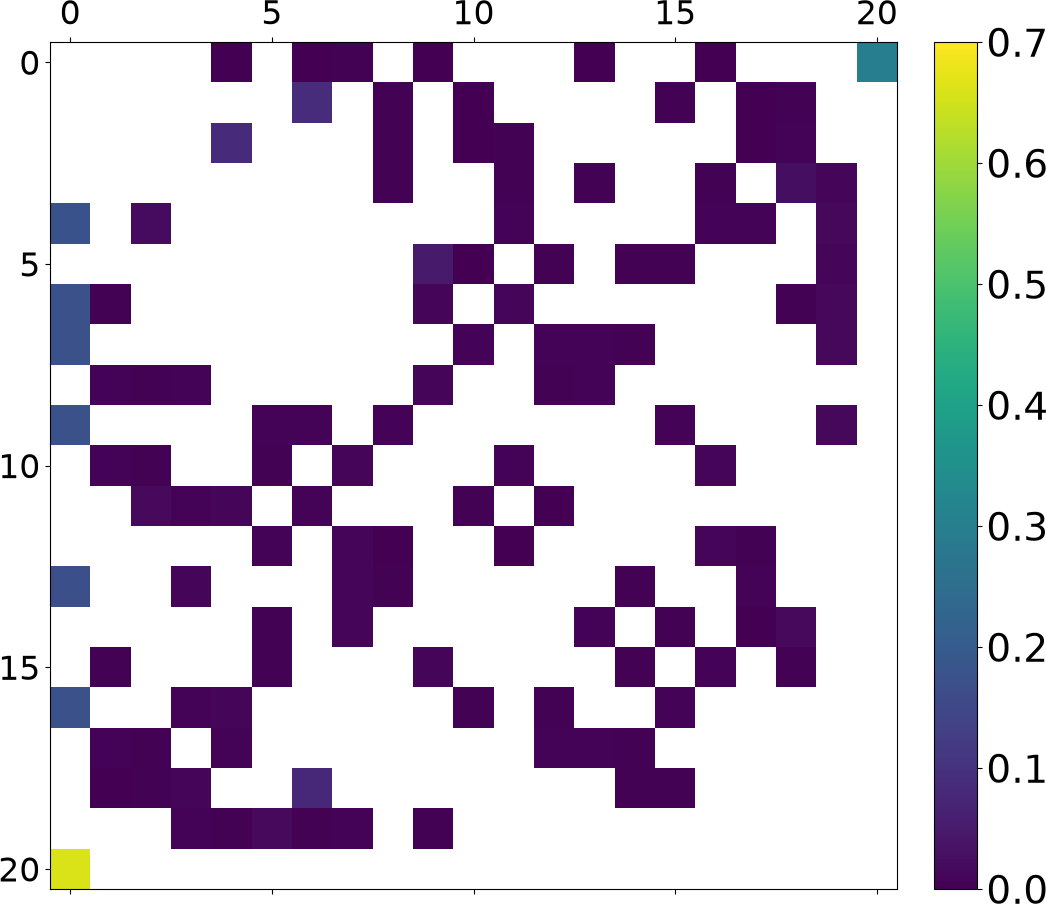}\\
\multicolumn{2}{c|}{(c)}& (i)& (ii)\\[2pt]
\hline
\rotatebox{90}{\qquad$\alpha=5$}&\rotatebox{90}{\qquad$\sigma=0.031$}&\rule{0pt}{3pt} \includegraphics[width=3.5cm]{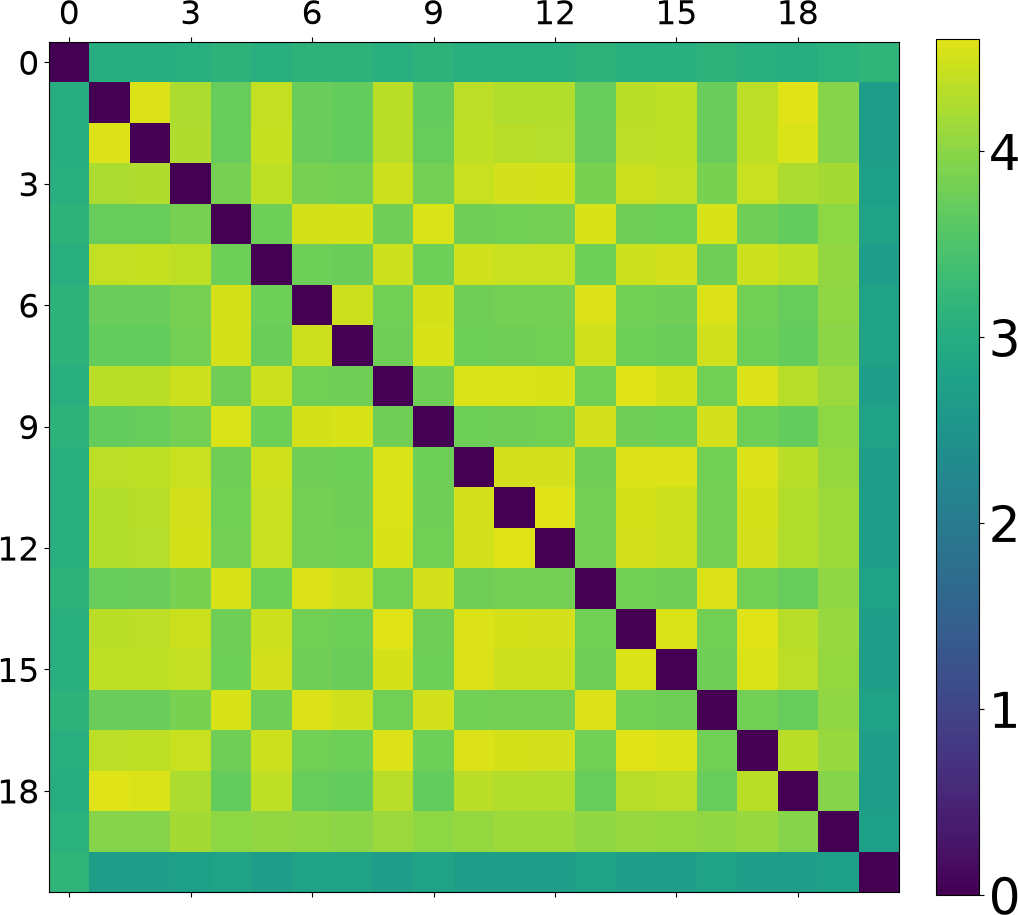}&\rule{0pt}{3pt} \includegraphics[width=3.5cm]{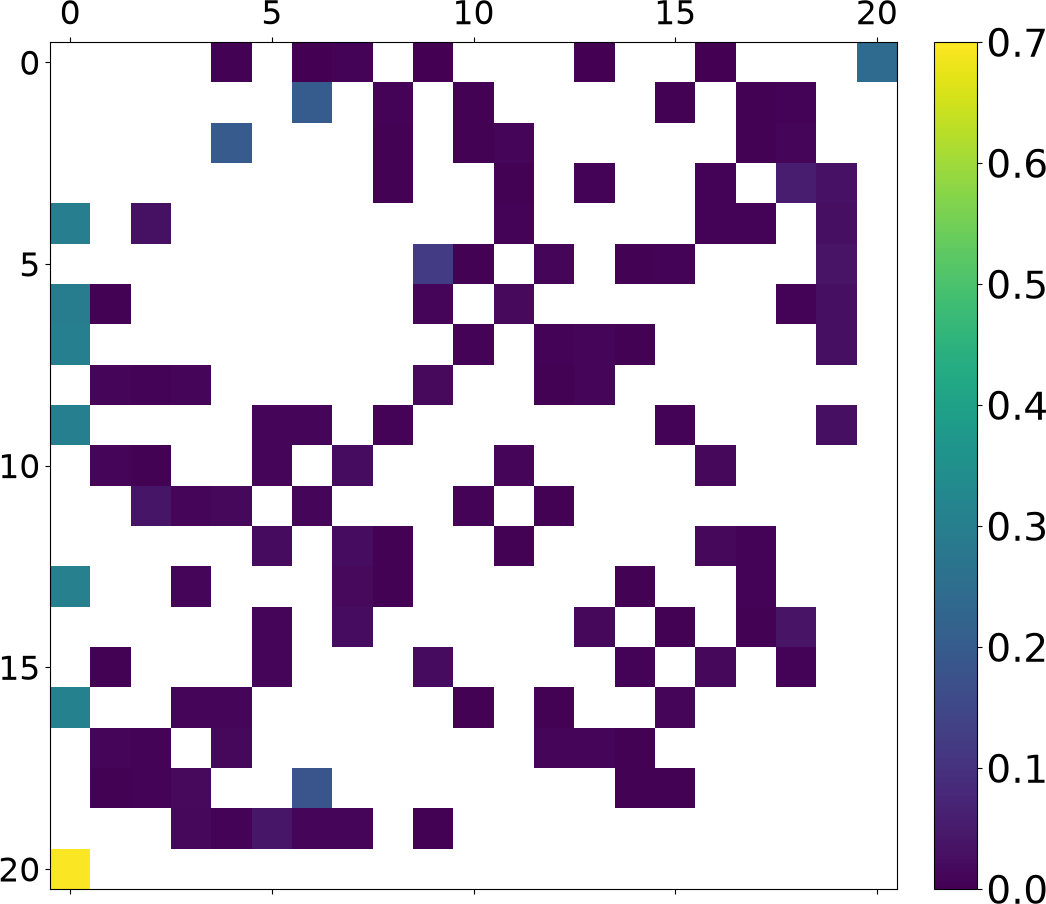}\\
\multicolumn{2}{c|}{(d)}& (i)& (ii)\\[2pt]
\hline
\end{tabular}
\caption{R\"{o}ssler dynamics coupled over the Ramanujan graph with the appended pendant vertex (as shown in Fig.~\ref{fig:alpha} (a)) with varying values of $\alpha$ and $\sigma$ as indicated in (a)-(d); for each set of parameters, we provide the (i) MI and (ii) CSE-inspired network flow heatmap as estimated using the $I_pi$ estimator for GPMI with the measure $m(f,g)=f\cdot g$ for $T=10,000$.\label{fig:Ram_MI}}
\end{figure}

If the synchronization ratio of every cut was plotted in this case (as was done in Fig.~\ref{fig:Xi_paths}), we would see a collection of constant values (e.g. the orange curve) that were very close together, and a single curve (similar to the red curve in Fig.~\ref{fig:Xi_paths}) would intersect that collection around $\alpha\approx 1.15$.  Although we are unable to say that, in fact, the pendant is no longer the driving disruption to complete synchrony after this tipping point, our measure does seem to capture a shift in noticeable bottleneck impacts. So, whether it determines the true limiting bottleneck or not, this measure does provide insight that is based on locally available information, which may be used in decentralized strategies for adjusting local weights to improve overall synchronization stability.

\section{Conclusions}
\label{sec:Conc}
Despite the deep connections that exist between spectral graph theory and the process of synchronization in network coupled dynamical systems, including the Master Stability Function viewpoint and the well-known Cheeger inequalities; neither the traditional isoperimetric number nor the more robust bottleneck measure of graph conductance are sufficient to characterize complete synchronization stability.  We have proposed a new graph property that we call the synchronization bottleneck measure, which is the global minimizer of a synchronization bottleneck ratio.  The associated cut of this measure does, in fact, identify small-set bottlenecks, and for parameterized pendant graphs, its value grows in a similar way to the algebraic connectivity until a transition point that can indicate a larger cut has become the true limiting bottleneck. However, due to the complexity of synchronization dynamics and the local nature of the definition of this property, it is important to note that it is not sufficient to fully characterize complete synchronization stability, especially when the Fiedler subspace is high dimensional or the graph conductance is small, making the associated cut a significant contributing factor to the failure to synchronize.  

Although this synchronization bottleneck measure may not completely characterize the limiting bottleneck, it provides a good approximation, and importantly, the definition of the ratio provides an excellent local measure that is computable through peer-to-peer (P2P) communication networks of distributed systems.  Whenever global knowledge of the network structure is not available, there are still local adaptations that can be made, especially in man-made systems, based on the limited available knowledge that can be gained from direct neighboring nodes to improve the synchronization stability of the overall system.  Future work will explore decentralized strategies for iterative adaptation to achieve globally optimal network structure. For example, each node might strive to not have any of their edges used in the cut that defines the overall synchronization bottleneck.  

In the case of diffusion-based clock synchronization over a P2P network, if a node has information related to the degrees of its neighbors (i.e. their current neighbor weighting), it should adjust the weighting used in its own distributed clock update to more closely align with higher degree nodes.  Increasing the edge weight to these higher degree nodes will increase the synchronization bottleneck ratio for any cut that included that link.  This increase in weight will follow a similar curve to the $\alpha$-pendant plots for $\Xi_G$; the goal being to reach a transition point where some other cut set becomes the minimizer and thus the limiting bottleneck of the system.  

Of course, this kind of adhoc adjustment would result in a directed network structure and additional algorithmic adjustments would be required to converge to a stable symmetrically weighted network structure.  Such an algorithm would necessarily need to prevent runaway edge weights as that would lead to so-called \textit{domination by neighbors}, wherein the internal dynamics of a hub node can become overwhelmed in traditional combinatorial Laplacian coupling schemes.  If sufficient path redundancy is not present among a hub's neighbors (also locally available information), such systems can experience \textit{oscillation death} from conflicting influences reducing the dynamics to a mean field. 

In some engineered domains, e.g. P2P communication networks, it is common practice to simply limit the number of allowable neighbors, and while this heuristic can prevent domination by neighbors, it also drastically alters the underlying topology of the network, altogether preventing the formation of hub nodes, which are critical to scale-free topologies. While hub-driven heterogeneity can be a challenge to synchronization, coupling through the normalized symmetric Laplacian can address this challenge in an alternative way, but this too can drastically change the system behavior, increasing convergence times.  These competing challenges drive the need for new algorithmic approaches to decentralized optimization.  We provide this definition as a step in the development of a more complete understanding of network coupled synchronization dynamics.

\section*{Acknowledgements}
The authors would like to dedicate this work to the memory of Erik Bollt for his many contributions to this effort prior to his passing. CTD gratefully acknowledges funding from the Air Force Office of Scientific Research under award no. 23RICOR001, and both CTD and JF gratefully acknowledge funding by the Army Research Office under award no. W911NF2310393.  Some of the computing for this project was performed on the ACRES cluster. We would like to thank Clarkson University and the Office of Information Technology for providing computational resources and support that contributed to these research results. Additional computational resources for this grant were provided by the National Science Foundation under Grant No. 1925596.

\bibliographystyle{unsrt}
\bibliography{xi}

\end{document}